\documentclass[12pt]{article}
\usepackage{graphicx}
\def\hybrid{\topmargin 0pt      \oddsidemargin 0pt
        \headheight 0pt \headsep 0pt
        \voffset=-0.5cm
        \textwidth 6.25in       
        \textheight 9.5in       
        \marginparwidth 0.0in
        \parskip 5pt plus 1pt   \jot = 1.5ex}
\catcode`\@=11
\def\marginnote#1{}

\newcount\hour
\newcount\minute
\newtoks\amorpm
\hour=\time\divide\hour by60
\minute=\time{\multiply\hour by60 \global\advance\minute by-\hour}
\edef\standardtime{{\ifnum\hour<12 \global\amorpm={am}%
        \else\global\amorpm={pm}\advance\hour by-12 \fi
        \ifnum\hour=0 \hour=12 \fi
        \number\hour:\ifnum\minute<10 0\fi\number\minute\the\amorpm}}
\edef\militarytime{\number\hour:\ifnum\minute<10 0\fi\number\minute}

\def\draftlabel#1{{\@bsphack\if@filesw {\let\thepage\relax
   \xdef\@gtempa{\write\@auxout{\string
      \newlabel{#1}{{\@currentlabel}{\thepage}}}}}\@gtempa
   \if@nobreak \ifvmode\nobreak\fi\fi\fi\@esphack}
        \gdef\@eqnlabel{#1}}
\def\@eqnlabel{}
\def\@vacuum{}
\def\draftmarginnote#1{\marginpar{\raggedright\scriptsize\tt#1}}
\def\draftlabel#1{{\@bsphack\if@filesw {\let\thepage\relax
   \xdef\@gtempa{\write\@auxout{\string
      \newlabel{#1}{{\@currentlabel}{\thepage}}}}}\@gtempa
   \if@nobreak \ifvmode\nobreak\fi\fi\fi\@esphack}
        \gdef\@eqnlabel{#1}}
\def\@eqnlabel{}
\def\@vacuum{}
\def\draftmarginnote#1{\marginpar{\raggedright\scriptsize\tt#1}}

\def\draft{\oddsidemargin -.5truein
        \def\@oddfoot{\sl preliminary draft \hfil
        \rm\thepage\hfil\sl\today\quad\militarytime}
        \let\@evenfoot\@oddfoot \overfullrule 3pt
        \let\label=\draftlabel
        \let\marginnote=\draftmarginnote
   \def\@eqnnum{(\theequation)\rlap{\kern\marginparsep\tt\@eqnlabel}%
\global\let\@eqnlabel\@vacuum}  }

\def\numberbysection{\@addtoreset{equation}{section}
        \def\theequation{\thesection.\arabic{equation}}}

\def\underline#1{\relax\ifmmode\@@underline#1\else
        $\@@underline{\hbox{#1}}$\relax\fi}

\def\titlepage{\@restonecolfalse\if@twocolumn\@restonecoltrue\onecolumn
     \else \newpage \fi \thispagestyle{empty}\c@page\z@
        \def\thefootnote{\fnsymbol{footnote}} }

\def\endtitlepage{\if@restonecol\twocolumn \else  \fi
        \def\thefootnote{\arabic{footnote}}
        \setcounter{footnote}{0}}  
\relax

\numberbysection
\hybrid

\newfont{\Bbb}{msbm10 scaled 1\@ptsize00}
\newfont{\Bbbb}{msbm7 scaled 1\@ptsize00}

\newcommand{\DDD}{\raise-1pt\hbox{$\mbox{\Bbbb D}$}}

\newcommand{\PP}{\mbox{\Bbb P}}        

\newcommand{\UUU}{\raise-1pt\hbox{$\mbox{\Bbbb U}$}}

\newcommand{\ZZ}{\mbox{\Bbb Z}}
\newcommand{\z}{\raise-1pt\hbox{$\mbox{\Bbbb Z}$}}

\newcommand{\vf}{\varphi}

\newcommand{\om}{\omega}
\newcommand{\vth}{\vartheta_1}
\newtheorem{predl}{Proposition}

\def\beq{\begin{equation}}
\def\eeq{\end{equation}}
\def\p{\partial}

\def\square{\hfill
{\vrule height6pt width6pt depth1pt} \break \vspace{.01cm}}

\begin{document}

\begin{titlepage}
\setcounter{page}{1}

\title{Quantum Painlev\'e-Calogero Correspondence}

\author{A. Zabrodin
\thanks{Institute of Biochemical Physics,
Kosygina str. 4, 119991 Moscow, Russia
and ITEP, Bol. Cheremushkinskaya str. 25, 117259 Moscow, Russia,
E-mail: zabrodin@itep.ru}
\and
A. Zotov
\thanks{ITEP, Bol. Cheremushkinskaya str. 25, 117259 Moscow, Russia,
E-mail: zotov@itep.ru}}

\date{July 2011}

\maketitle

\vspace{-7cm} \centerline{ \hfill ITEP-TH-23/11} \vspace{7cm}

\begin{abstract}

The Painlev\'e-Calogero correspondence is extended to
auxiliary linear problems associated with Painlev\'e equations.
The linear problems are represented in a new form
which has a suggestive interpretation as a
``quantized'' version of the Painlev\'e-Calogero correspondence.
Namely, the linear problem responsible for the
time evolution is brought into the form of non-stationary
Schr\"odinger equation in imaginary time,
$\p_t \psi =(\frac{1}{2}\, \p_x^2 +V(x,t))\psi$,
whose Hamiltonian is a natural quantization of the
classical Calogero-like Hamiltonian $H=\frac{1}{2}\, p^2 +V(x,t)$
for the corresponding
Painlev\'e equation.

\end{abstract}

\tableofcontents


\end{titlepage}

\section{Introduction}

The famous six nonlinear ordinary second-order differential
equations discovered by P.Painlev\'e, R.Fuchs and B.Gambier
\cite{Painleve1,Fuchs,Gambier} in the beginning
of the ${\rm XX}$ century are nowadays known as
the Painlev\'e equations ${\rm I}$--${\rm VI}$
(${\rm P}_{\rm I}$--${\rm P}_{\rm VI}$).
Since that time they were extensively studied
and they still remain to be among the most important and
most interesting differential equations in mathematics and
mathematical physics \cite{book,book1}. Their applications
include self-similar reductions of non-linear integrable
partial differential equations \cite{FN},
correlation functions of integrable models \cite{Bar,JMMS},
quantum gravity and string theory \cite{qg}, topological field
theories \cite{Dubrovin}, 2D polymers \cite{Zamolodchikov},
random matrices \cite{TW,FW} and stochastic
growth processes \cite{LTW}, to mention only few
applications and few references.

The idea to associate a system of {\it linear} differential
equations with each Painlev\'e equation goes back to
the seminal work by R.Fuchs \cite{Fuchs}.
In fact the theory of Painlev\'e equations is intrinsically
related to the monodromy properties of linear ordinary
differential equations with rational coefficients.
Remarkably, the equations from the Painlev\'e list
describe monodromy preserving deformations of linear
differential equations with essential singularities.
The classical references on the subject are \cite{Garnier,Schlesinger}.
The monodromy approach
was further developed by H.Flaschka and A.Newell \cite{FN}
and by M.Jimbo, T.Miwa and K.Ueno
in the series of works \cite{JM81-1,JM81-2,JM81-3}, see also
book \cite{IN}.
At present different types of linear problems
(scalar \cite{Fuchs,Garnier}, 2$\times$2-matrix
\cite{JM81-2} or 3$\times$3-matrix \cite{JKT07})
are known to be associated with Painlev\'e equations.

The Hamiltonian theory of the Painlev\'e equations
is dated back to the work \cite{Malmquist} (for the
modern developments and the extension to general Schlesinger
systems see \cite{DM07}).
It turns out that all the six equations have
a Hamiltonian structure with time-dependent
Hamiltonian functions which are polynomials
in the dependent variable (the coordinate) and suitably chosen
conjugate momentum. They are referred to
as Okamoto's Hamiltonians \cite{Okamoto}. However,
the Okamoto's Hamiltonians
for ${\rm P}_{\rm II}$--${\rm P}_{\rm VI}$
equations are of a more complicated form than just
momentum squared plus potential. This makes a direct
interpretation of Painlev\'e equations as classical
mechanical systems (a point-like particle on the line moving
in a time-dependent potential) problematic. Nevertheless, such an
interpretation appears to be possible after a non-trivial
canonical transformation which accomplishes the
{\it Painlev\'e-Calogero correspondence}.

The phenomenon known in the literature as the
(classical) Painlev\'e-Calogero
correspondence \cite{LO97} consists in the possibility to represent,
by means of explicitly known transformations of the dependent and
independent variables, all the six Painlev\'e equations as
non-autonomous Hamiltonian systems
$$
\p_t x =\frac{\p H}{\p p}\,, \quad
\quad \p_t p =-\frac{\p H}{\p x}
$$
with the standard one-particle Hamiltonian of the
canonical form
$H=p^2 /2 +V(x,t)$ for some potential $V(x,t)$ which
explicitly depends on time $t$. In the case of
${\rm P}_{\rm VI}$ this Hamiltonian system
resembles the elliptic Calogero
model with 2 particles in the center of mass coordinates,
whence the name Painlev\'e-Calogero correspondence.
(To be more precise, the ${\rm P}_{\rm VI}$ equation
is a non-autonomous version of a special rank-one case of
the Inozemtsev's extension \cite{Inoz} of the elliptic Calogero model.)
For the ${\rm P}_{\rm VI}$ equation this remarkable observation
was made by Yu.Manin \cite{Manin98} who revived the almost forgotten
work by Painlev\'e himself \cite{Painleve1906}. Later,
K.Takasaki \cite{Takasaki01} extended this result to the other
equations from the Painlev\'e list. In principle, this extension
can be achieved by a special degeneration process from ${\rm P}_{\rm VI}$
to the lower members of the Painlev\'e family. Although the
resulting Hamiltonian systems hardly resemble any Calogero-like
models, the name ``Painlev\'e-Calogero correspondence''
has been extended to these cases as well. This also suggests
generalizations to higher rank systems which were studied in
\cite{Takasaki01}.

The explicit form of the canonical transformations from the
Okamoto's Hamiltonian systems to Calogero-like ones was found
in \cite{Takasaki01}. Here we need only the coordinate part of this
transformation which is described by the following theorem.

\noindent
\paragraph{Theorem 1 \cite{Takasaki01}.}{\it
For any of the six equations
from the Painlev\'e list written
for a variable $y(T)$ there exists a change of variables
$(y,T)\rightarrow (u,t)$ of the form $y=y(u,t)$, $T=T(t)$
that maps the Painlev\'e equation to a second-order differential
equation of the form
\beq\label{d000}
\ddot u = -\p_u V(u,t)
\eeq
which is equivalent to a non-autonomous Hamiltonian system
$\dot u =\p H(p, u,t)/\p p$, $\dot p=-\p H(p,u,t)/\p u$
with the Hamiltonian
\beq\label{d000a}
H(p, u,t)=\frac{p^2}{2} +V(u,t)
\eeq
where $V(u,t)$ is a time-dependent potential written in terms
of rational, hyperbolic or elliptic functions.
}

\noindent
This statement was proved in \cite{Takasaki01} by
giving explicit formulas for the corresponding changes
of variables (see the table below).
We call (\ref{d000}) the {\it Calogero form} of the
Painlev\'e equation.

The aim of this paper is to extend the Painlev\'e-Calogero correspondence
to the linear problems associated with the Painlev\'e equations.
In fact we suggest a new form of the linear problems
which allows us to interpret it as a ``quantized'' version of the
Painlev\'e-Calogero correspondence. In other words,
linearization,
i.e., going to the associated linear problems, appears to be
equivalent to quantization of the Painlev\'e equations
regarded as classical mechanical systems.

The starting point is a system of two first-order linear
partial differential equations (PDE) in two variables
for a 2-component vector-function $(\psi _1, \psi_2)^{\rm t}$
of the form presented, for example, in \cite{JM81-2}.
The two variables are
the spectral parameter and the deformation parameter.
As is well known,
compatibility of the system is equivalent to the zero
curvature condition for the connection represented
by $2\times 2$ matrices depending on the two variables.
The next step is the change of the dependent and independent
variables that leads to the Calogero-like form of the
Painlev\'e equations, supplemented by a suitable change of the
spectral parameter (polynomial for
${\rm P}_{\rm I}$, ${\rm P}_{\rm II}$, ${\rm P}_{\rm IV}$,
exponential for ${\rm P}_{\rm III}$, hyperbolic for ${\rm P}_{\rm V}$
and elliptic for ${\rm P}_{\rm VI}$).
At this step the spectral parameter and the deformation parameter
acquire the meaning of the coordinate and time variables for a
non-autonomous dynamical system with one degree of freedom.
After an additional
diagonal gauge transformation of a special form,
the linear problems transformed in this way
should be rewritten as a pair of two compatible linear PDE's
for a scalar $\psi$-function
$\psi =\psi_1$ (the first component of the vector function).
One of them is an ordinary second-order differential equation
with coefficients explicitly depending on time and on the dependent
variable. After a simple transformation of the $\psi$-function
the term with the first order derivative cancels, and one obtains
a stationary Schr\"odinger equation with a potential function which
depends on time in both explicit and implicit ways, with the implicit
dependence coming from the dependent variable.
The isomonodromy problem for this equation, i.e.,
time-dependent deformation of the potential preserving the
monodromy of solutions, is known to be
equivalent to the Painlev\'e equation.

The key new element introduced
in this paper is the second equation of the pair, the one
describing the time evolution. We show that for all the six
Painlev\'e equations (and any values of the
standard parameters $\alpha , \beta , \gamma , \delta$
involved)
it can be represented in the form of the
{\it non-stationary Schr\"odinger equation} in imaginary time,
$$
\p_t \Psi = \hat H \Psi \,,
$$
whose Hamiltonian is the standard 1D Schr\"odinger operator $\hat
H=\frac{1}{2}\, \p_{x}^{2}+V(x, t)$ which is a {\it natural
quantization of the classical Calogero-like Hamiltonian} associated
with the Painlev\'e equation at hand (to be more precise, for ${\rm
P}_{\rm VI}$ the parameters $\alpha , \beta , \gamma , \delta$ in
the quantized Hamiltonian appear to be shifted by ``quantum
corrections'' $\pm \frac{1}{8}$, similar shifts of some of the
parameters take place also for ${\rm P}_{\rm V}$ and ${\rm P}_{\rm
IV}$). Therein lies the quantum Painlev\'e-Calogero correspondence,
or a classical-quantum correspondence for the Painlev\'e equations.
Indeed, on the Calogero side, one now has a {\it quantum}
Calogero-like or Inozemtsev model in a non-stationary state
described by the wave function $\Psi$
which differs from $\psi$ by a coordinate-independent factor.
On the Painlev\'e side, this
$\Psi$-function is a common solution to the linear problems
associated with the Painlev\'e equation. Solutions of the Painlev\'e
equation itself can be extracted from the asymptotic behavior of the
$\Psi$-function near singular points. The main results
of this work are
summarized in the following ``quantum'' version of Theorem 1:

\noindent
\paragraph{Theorem 2.}{\it  For any of the six equations
from the Painlev\'e list written in the Calogero form (\ref{d000})
as classical Hamiltonian systems with time-dependent Hamiltonians
$H(p,u,t)$ (\ref{d000a})
there exists a pair of compatible linear problems
\beq\label{d001}
\left \{ \begin{array}{l} \p_{x}\mathbf{\Psi} ={\bf
U} (x,t,u,\dot{u},\{c_k\})\mathbf{\Psi}
\\ \\
\p_{t}\mathbf{\Psi} ={\bf V} (x,t,u,\dot{u},\{c_k\})\mathbf{\Psi}
\end{array}\right. \,,
\quad \quad \mathbf{\Psi} =\left (\begin{array}{l} \psi_1 \\
\psi_{2} \end{array}\right ),
\eeq
where ${\bf U}$ and ${\bf V}$ are
$sl_2$-valued functions, $x$ is a spectral parameter,
$t$ is the time variable and $\{c_k\}=\{\alpha , \beta ,
\gamma , \delta \}$ is the set of parameters involved
in the Painlev\'e equation, such that
\begin{itemize}
\item[1)]
The zero curvature condition
\beq\label{d002}
\p_t{\bf U}-\p_x {\bf V}+[{\bf U},{\bf V}]=0
\eeq
is equivalent to the Painlev\'e equation (\ref{d000})
for the variable $u$ defined as any (simple) zero of the
right upper element of the matrix ${\bf U}(x,t)$ in the spectral parameter:
${\bf U}_{12}(u, t)=0$;
\item[2)]
The function $\Psi =e^{\int^t H(\dot u, u,t')dt'}\psi _1$
where $\psi_1$ is the
first component of $\mathbf{\Psi}$ satisfies the non-stationary
Schr\"odinger equation in imaginary time
\beq\label{d003}
\p_t\Psi =\left(\frac{1}{2}\, \p_x^2 +\tilde V (x,t) \right)\Psi
\eeq
with the potential
$$
\tilde V(x,t)=V(x,t, \{\tilde c_k\})=\frac{1}{2}
\Bigl [\det({\bf U})-\p_x {\bf
U}_{11}+2{\bf V}_{11}\Bigr ]
$$
which coincides with the classical
potential $V(x,t)=V(x,t, \{c_k\})$
up to possible shifts of the parameters $\{c_k\}$:
$$
\begin{array}{l}
(\tilde \alpha , \tilde \beta  ) =(\alpha ,
\beta +\frac{1}{2}) \quad \mbox{for ${\rm P}_{\rm IV}$},
\\ \\
(\tilde \alpha ,
\tilde \beta , \tilde \gamma , \tilde \delta )=
(\alpha -\frac{1}{8}, \beta +\frac{1}{8}, \gamma , \delta )
\quad \mbox{for ${\rm P}_{\rm V}$}
\\ \\
(\tilde \alpha ,
\tilde \beta , \tilde \gamma , \tilde \delta )=
(\alpha -\frac{1}{8}, \beta +\frac{1}{8},
\gamma -\frac{1}{8}, \delta +\frac{1}{8}) \quad
\mbox{for ${\rm P}_{\rm VI}$}.
\end{array}
$$
\end{itemize}
}

For reader's convenience we collect the changes of variables
from the original $y,T$ to $u,t$ required for passing to the
Calogero form and the corresponding change
of the spectral parameter from rational one, $X$, to
$x$, in the following table:

\begin{center}
\begin{tabular}{|c|c|c|c|c|}
\hline
$\displaystyle{\phantom{\int}}$
Equation $\displaystyle{\phantom{\int_{\int}^{\int}}}$
& $y(u,t)$ & $T(t)$ & $X(x,t)$ & ${\bf U}_{12}(x,t)$ \\
\hline
$\displaystyle{\phantom{\int}}$ ${\rm P}_{\rm I}$
& $u$ & $t$ & $x$ &$x-u$\\
\hline
$\displaystyle{\phantom{\int}}$ ${\rm P}_{\rm II}$
& $u$ & $t$ & $x$ &$x-u$\\
\hline
$\displaystyle{\phantom{\int}}$ ${\rm P}_{\rm IV}$
& $u^2$ & $t$ & $x^2$ &$x^2 -u^2$ \\
\hline
$\displaystyle{\phantom{\int}}$ ${\rm P}_{\rm III}$
& $e^{2u}$ & $e^t$ & $e^{2x}$ & $2e^{t/2}\sinh (x-u)$ \\
\hline
$\displaystyle{\phantom{\int}}$ ${\rm P}_{\rm V}$
& $\coth ^2 u$ & $e^{2t}$ & $\cosh ^2 x$ & $2e^{t}
\sinh (x\! -\! u)\sinh (x\! +\! u)$ \\
\hline
$\displaystyle{\phantom{\int^{A}_{A}}} {\rm P}_{\rm VI}$
& $
\frac{\wp (u)-\wp (\omega _1)}{\wp (\omega_2)-\wp (\omega_1)}$
& $
\frac{\wp (\omega _3)-\wp (\omega _1)}{\wp (\omega_2)-\wp (\omega_1)}$
& $
\frac{\wp (x)-\wp (\omega _1)}{\wp (\omega_2)-\wp (\omega_1)} $ &
$
\vartheta _1(x-u)\vartheta _1(x+u) h(u,t) $ \\
\hline
\end{tabular}
\end{center}

\noindent
In the last column the right upper element of the matrix
${\bf U}(x,t)$ is given. One can see that in all cases
$u$ is indeed
a simple zero of ${\bf U}_{12}(x,t)$.
The function $h(u,t)$ is some function
of $u,t$ only to be specified in Section 8.
The Weierstrass $\wp$-function
$\wp (z)=\wp (z|1, \tau )$ and the Jacobi theta-function
$\vartheta_1(x)=\vartheta_1(x|\tau )$ in the last
line of the table depend on $t$ in a non-trivial way
through the second period $\tau = 2\pi i t$. The half-periods
are defined as $\omega _1 = \frac{1}{2}$, $\omega_2 =
\frac{1}{2}(1+\tau )$, $\omega _3 =\frac{1}{2}\, \tau$.

When this work was completed, we were informed by B.Suleimanov
that he realized the role of non-stationary Schr\"odinger-like
equation in linear problems for Painlev\'e equations
back in 1994 and obtained similar results \cite{Suleimanov94}.
In distinction to our approach, he starts
with the scalar linear problems of the Fuchs-Garnier type
with rational spectral parameter \cite{Fuchs,Garnier}
and shows that their compatibility implies yet another linear equation
for the same wave function, which is of the
non-stationary Schr\"odinger form, with quantum Hamiltonian being
a quantization of the corresponding Okamoto's Hamiltonian.
The precise connection between the two approaches deserves
further elucidation.

The presentation is organized in such a way that each Painlev\'e equation
is discussed in a separate section, in the order of increasing complexity,
from ${\rm P}_{\rm I}$ to ${\rm P}_{\rm VI}$ (Sections 3--8).
We tried to make each section
self-contained, so that they could be read independently of each other.
However, each section contains references to Section 2, where the general
construction is outlined.
Note that in our list
${\rm P}_{\rm IV}$ stands before ${\rm P}_{\rm III}$ because in
a certain sense
the complexity of the latter exceeds that of the former.
This is due to the fact that
the ${\rm P}_{\rm I}$, ${\rm P}_{\rm II}$ and ${\rm P}_{\rm IV}$ equations
need rational parametrization to be represented in the Calogero-like form
while ${\rm P}_{\rm III}$ and ${\rm P}_{\rm V}$ require
exponential
and hyperbolic parametrizations for that purpose.
The highest member, ${\rm P}_{\rm VI}$, is the most complicated
object. It requires parametrization in terms of elliptic functions.
One can see that
the calculations which are necessary to prove
Theorem 2 and to verify the classical-quantum
correspondence, being really short and transparent for ${\rm P}_{\rm I}$,
become very long and tedious for ${\rm P}_{\rm VI}$. In the case
of ${\rm P}_{\rm VI}$ (and to some extent of ${\rm P}_{\rm V}$),
the situation is aggravated by the fact that neither
the change of the spectral parameter nor the gauge transformation are
known from the very beginning and should be either guessed
or found by solving a differential equation.
The three appendices are all related to
the ${\rm P}_{\rm VI}$ equation. In Appendix A some details
of explicit verification of the zero curvature
condition are given. Appendix B contains the necessary
information on theta-functions and elliptic functions.
In Appendix C the special diagonal gauge transformation
together with the change of the spectral parameter for the
linear problems for the ${\rm P}_{\rm VI}$ equation is derived.

\section{The general scheme}

\subsection{Linear problems and compatibility conditions}

As is known, any Painlev\'e equation I-VI
can be represented as the compatibility condition for
a pair of linear problems depending on a spectral parameter.
We need the linear problems such that they lead
directly to the Painlev\'e equations in the Calogero form.
They can be
obtained from the linear problems with rational spectral
parameter by a proper change of variables.
The existence of such a change of variables will be
proved separately for each equation
${\rm P}_{\rm I}$-${\rm P}_{\rm VI}$ by an explicit
calculation. Now suppose that we are given with such
a pair of linear problems:
\beq\label{d1}
\left \{ \begin{array}{l}
\p_{x}\mathbf{\Psi} ={\bf U} (x,t)\mathbf{\Psi}
\\ \\
\p_{t}\mathbf{\Psi} ={\bf V} (x,t)\mathbf{\Psi}
\end{array}\right. \,,
\quad \quad
\mathbf{\Psi} =\left (\begin{array}{l}
\psi_1 \\ \psi_{2} \end{array}\right ),
\eeq
where the 2$\times$2 matrices ${\bf U},{\bf V}$ explicitly
depend on the spectral
parameter $x$ (which in our approach
has the meaning of coordinate), on the deformation parameter
$t$ (which in our approach has the meaning of time)
and contain an unknown functions of $t$ to be constrained
by the condition that the two equations have a family of common
solutions.
This function
is going to satisfy one of the six Painlev\'e equations (in the Calogero
form).
In fact the latter is equivalent
to the compatibility of the linear problems expressed as
the zero curvature equation (integrability condition)
\beq\label{P1c}
\p_x {\bf V} -\p_t {\bf U} +[{\bf V}, {\bf U}]=0\,.
\eeq
Set
$$
{\bf U}=\left ( \begin{array}{cc}
a & b\\  c & d
\end{array}\right ),
\quad \quad
{\bf V}=\left ( \begin{array}{cc}
A & B\\  C & D
\end{array}\right ).
$$
Our matrices  ${\bf U},{\bf V}$
will be always traceless, i.e., $a+d=0$, $A+D=0$.
In this notation,
the zero curvature equation yields:
\beq\label{zc1}
\left \{ \begin{array}{l}
\displaystyle{a_t -A_x +b C - c B =0}
\\ \\
\displaystyle{b_t -B_x +2aB -2bA=0}
\\ \\
\displaystyle{c_t -C_x +2cA -2a C=0\,.}
\end{array}
\right.
\eeq
Here and below $a_t$, $A_x$, etc mean partial derivatives
with respect to $t,x$. To avoid a misunderstanding, we
emphasize that the time variable $t$ enters the matrix elements
in two ways: explicit and implicit. The latter means
the time dependence through the
unknown functions of $t$ (dependent variables). The notation
$a_t$, etc implies the full time differentiating which takes into
account the time dependence of the both types.

The function that satisfies the Painlev\'e equation
in the Calogero form will be denoted by $u=u(t)$.
It can be defined as zero
of the right upper element of the matrix ${\bf U}(x,t)$
as a function of the spectral parameter $x$:
$b(u)=0$. We will see that this zero is always of the
first order and different possible choices (in the case when
the function $b(x)$ has more than one zero in a suitably
chosen fundamental domain) lead to the same equation.

It is important that the matrix functions
${\bf U}(x,t),{\bf V}(x,t)$
have poles in $x$ at the points which may depend on time
{\it but not through the dependent variable $u$}.
In fact for ${\rm P}_{\rm I}$ -- ${\rm P}_{\rm V}$ equations
they are time independent while
for the ${\rm P}_{\rm VI}$ equation two poles are fixed and
other two linearly depend on the time variable.

In what follows we will choose the matrices
${\bf U}$, ${\bf V}$ such that
\beq\label{Bb}
b_x =2B.
\eeq
(the meaning and advantages of this condition will be clear later).
Given any two matrix functions
${\bf U}$, ${\bf V}$,
this equality can be always attained by means of
a suitable diagonal
gauge transformation of the linear system (\ref{d1})
(see below).
In principle, one can then exclude $A$ and $C$ from the
zero curvature equations
(\ref{zc1}) and obtain
a functional relation for $a,b$ and $c$ but we will not follow this route here.
Let us only mention, for future reference,
that if the zero curvature equation and the condition (\ref{Bb}) are
imposed, then
$A$ is expressed through $a$ and $b$ as follows:
\beq\label{Aab}
2A = \frac{b_t +ab_x}{b}-\frac{b_{xx}}{2b}\,.
\eeq

The system (\ref{d1}) admits gauge transformations
$
\mathbf{ \tilde \Psi}  =\Omega  \mathbf{\Psi}
$
with a matrix $\Omega$ which can depend on
$x,t$. The gauge transformed system has the same form
\beq\label{d1a}
\left \{ \begin{array}{l}
\p_{x} \mathbf{ \tilde \Psi}={\bf \tilde U} (x,t)\mathbf{ \tilde \Psi}
\\ \\
\p_{t}\mathbf{ \tilde \Psi} ={\bf \tilde V} (x,t)\mathbf{ \tilde \Psi}
\end{array}\right.
\eeq
with
\beq\label{tilde}
{\bf \tilde U} =\Omega^{-1}{\bf U}\Omega -\Omega^{-1}\p_x \Omega \,,
\quad \quad
{\bf \tilde V} =\Omega^{-1}{\bf V}\Omega -\Omega^{-1}\p_t \Omega\,.
\eeq
In the next sections this transformation will be applied
in the opposite direction, from matrices ${\bf \tilde U}$,
${\bf \tilde V}$ obtained at an
intermediate stage of calculations to matrices ${\bf U}$, ${\bf V}$ in
the final form. This is equivalent to applying the inverse
transformation.
In particular, let
\beq\label{om}
\Omega =\left (\begin{array}{cl}
\omega & 0\\ 0&\omega^{-1}\end{array}\right )
\eeq
be a diagonal matrix, then
\beq\label{tilde1}
{\bf U}=\left ( \begin{array}{cc}
\tilde a +\p_x \log \omega & \tilde b\omega^2\\ \\
\tilde c\omega^{-2} & \tilde d -\p_x \log \omega
\end{array}\right ),
\quad \quad
{\bf V}=\left ( \begin{array}{cc}
\tilde A +\p_t \log \omega & \tilde B\omega^2\\  \\
\tilde C\omega^{-2} & \tilde D -\p_t \log \omega
\end{array}\right ).
\eeq

Let us consider the two linear problems (\ref{d1})
in detail. Explicitly, we have:
$$
\left \{ \begin{array}{l}
\p_x \psi _1 =a\psi _1 +b \psi_2
\\ \\
\p_x \psi _2 =c\psi_1 +d \psi_2
\end{array} \right. \,,
\quad \quad
\left \{ \begin{array}{l}
\p_t \psi _1 =A\psi _1 +B \psi_2
\\ \\
\p_t \psi _2 =C\psi_1 +D \psi_2
\end{array} \right.
$$
Applying $\p_x$ to the first equation of the first system,
and using the second equation, we obtain
\beq\label{P1d}
\p_{x}^{2}\psi_1 -(a+d)\p_x \psi_1 +
(ad-bc) \psi_1 -(a_x \psi_1 +b_x \psi_2)=0.
\eeq
Using the linear equations above, one can express $\psi_2$ through
$\psi_1$ in two different ways:
\beq\label{var}
\psi_2 =\frac{\p_x\psi_1 -a\psi_1}{b}=
\frac{\p_t\psi_1 -A\psi_1}{B}\,.
\eeq
The first possibility leads to a closed ordinary
second-order differential equation for $\psi_1$
while the second one leads to a partial
differential equation for $\psi_1$
as a function of $x,t$. As we shall see soon, both
have the form of Schr\"odinger equations, stationary and non-stationary.
This pair of scalar equations is equivalent to the original
system (\ref{d1}) in the sense that their compatibility
implies Painlev\'e equations for the dependent variable.
One can also say that
the second equation describes isomonodromic deformations of the
first one.
Let us consider them separately.
From now on we will write simply
$\psi$ instead of $\psi_1$.

\subsection{Ordinary second-order differential equation}

Using the first equality in (\ref{var}), we get an ordinary
second-order differential equation for $\psi :=\psi_1$:
$$
\p_{x}^{2}\psi -\Bigl (a+d+\frac{b_x}{b}\Bigr )\p_x \psi +
\Bigl (ad-bc -a_x +\frac{b_x a}{b}\Bigr )\psi =0.
$$
The coefficient functions here are expressed through
entries of the matrix ${\bf U}(x,t)$.
For traceless matrices with the condition (\ref{Bb}) the
equation acquires the form
$$
\p_{x}^{2}\psi -\frac{b_x}{b}\p_x \psi +
\Bigl (ad-bc -a_x +2A -\frac{b_t}{b}+\frac{b_{xx}}{2b}\Bigr )\psi =0
$$
(here for the transformation of the
last term (\ref{Aab}) has been used)
or
\beq\label{ode2}
\left (\frac{1}{2}\, \p_{x}^{2} -
\frac{b_x}{2b} \, \p_x \,  + W (x)\right )\psi =0,
\eeq
where
\beq\label{ode3}
W(x)=\frac{1}{2}(ad -bc -a_x +2A) -\frac{1}{2b}
\Bigl ( -\p_t +\frac{1}{2}\, \p_x^2\Bigr )b.
\eeq

The substitution $\psi = \sqrt{b}\, \check \psi$ kills the
first derivative term in eq. (\ref{ode2}) and brings it to the form
of stationary Schr\"odinger equation
\beq\label{ode2a}
\left (\frac{1}{2}\, \p_{x}^{2} +\check W (x)\right )\check \psi =0
\eeq
with the potential
\beq\label{checkW}
\check W= W +\frac{1}{4} \, \p_x^2 \log b -
\frac{1}{8}\, (\p_x \log b)^2.
\eeq
This equation has formal solutions with the WKB-like
asymptotes near poles of the potential:
\beq\label{ode4}
\check \psi (x) \cong\,\, (-2\check W)^{-\frac{1}{4}}\,
e^{\pm \int^x \sqrt{-2\check W}dx'}.
\eeq
An expansion of the right hand side near singularities of the potential
allows one to extract solutions to the corresponding
Painlev\'e equation.

\subsection{Non-stationary Schr\"odinger equation}

The second possibility in (\ref{var}) is more interesting for us here.
It leads to a partial differential equation for $\psi =\psi_1$
as a function of $x,t$:
\beq\label{P1e}
\p_{x}^{2}\psi -(a+d)\p_x \psi +
(ad-bc) \psi -\Bigl (a_x -\frac{b_x A}{B}\Bigr )\psi
-\frac{b_x}{B}\, \p_t \psi=0.
\eeq
The coefficient functions here are expressed through
entries of the both matrices ${\bf U}(x,t)$, ${\bf V}(x,t)$.
For traceless matrices with the condition (\ref{Bb})
the equation simplifies:
\beq\label{P1ee}
\p_{x}^{2}\psi  +
\Bigl (ad-bc -a_x +2A\Bigr )\psi
-2\p_t \psi =0.
\eeq
The role of the condition (\ref{Bb}) is thus to make constant
the coefficient in front of the time derivative
(the specific value $2$ of the constant is just a matter
of normalization).

Equation (\ref{P1ee}) is central for what follows.
Clearly, it has the form of a non-stationary Schr\"odinger
equation in imaginary time:
\beq\label{P1eea}
\p_t \psi =\left ( \frac{1}{2}\, \p_x^2 +U(x,t)\right )\psi
\eeq
with the potential
\beq\label{potU}
U(x,t)=\frac{1}{2}\,(ad-bc -a_x) +A \, =\,
\frac{1}{2}\, \det {\bf U}-\frac{a_x}{2}+A.
\eeq
In the subsequent sections 3--8 we verify, by means of the case study,
that for all Painlev\'e equations the dependent variable $u$
enters this potential only through
an irrelevant $x$-independent term while $x$-dependent
terms contain the time variable
in the explicit form only.
Moreover, this potential
turns out to be {\it the same as the classical mechanical
potential for Painlev\'e equations written in the Calogero form}.
(To be precise, we should point out that
for higher members of the Painlev\'e family,
${\rm P}_{\rm IV}$ -- ${\rm P}_{\rm VI}$, the
coefficients in front of different terms of the potential
may be modified).
This provides the quantum version of the
Painlev\'e-Calogero correspondence.

Summing up, we have reduced the linear system
(\ref{d1}) for the vector function
$\mathbf{\Psi}=(\psi_1 , \psi_2)^{\sf t}$ to two scalar
equations for $\psi :=\psi_1$:
\beq\label{scalareqs}
\left \{
\begin{array}{l}\displaystyle{
\left (\frac{1}{2}\, \p_{x}^{2} -\frac{1}{2}\,
(\p_x \log b) \, \p_x \,  + W (x,t)\right )\psi =0}
\\ \\
\displaystyle{
\p_t \psi =\left ( \frac{1}{2}\, \p_x^2 +U(x,t)\right )\psi \,.}
\end{array}
\right.
\eeq
The second equation describes isomonodromic deformations
of the first one and
their compatibility implies the Painlev\'e equation (in the Calogero
form) for the function
$u=u(t)$ defined as a (simple) zero of the function $b(x)$:
$b(u)=0$. The $x$-dependent part of the potential $U(x,t)$
does not contain the dependent variable $u$.
Note that the potentials $W$ and $U$ are related by
$$
W=U-\frac{1}{2}\, \p_t \log b +
\frac{1}{4}\, \p_{x}^2\log b+
\frac{1}{4}(\p_x \log b)^2,
$$
so the potential $W(x,t)$ has an apparent singularity at $x=u(t)$.

One can see that equations (\ref{scalareqs}) imply
the scalar linear problems in the form suggested by R.Fuchs \cite{Fuchs}
and R.Garnier \cite{Garnier}. Indeed, passing to the function
$\check \psi = \psi /\sqrt{b}$ and combining the two equations
(\ref{scalareqs}), one obtains the linear system
\beq\label{FG1}
\left \{
\begin{array}{l}\displaystyle{
\left (\frac{1}{2}\, \p_{x}^{2}  + \check W (x,t)\right )\check \psi =0}
\\ \\
\displaystyle{
\p_t \check \psi =
\left ( \Lambda \p_x - \frac{1}{2}\, (\p_x \Lambda ) \right )\check \psi }
\end{array}, \quad \quad
\Lambda := \frac{1}{2}\, \p_x \log b\,,
\right.
\eeq
with $\check W$ given by (\ref{checkW}), which is exactly of the
Fuchs-Garnier form. The integrability condition for this system is
\beq\label{FG2}
\p_t \check W = 2\check W \p_x \Lambda +\Lambda \p_x \check W +
\frac{1}{4}\, \p_x^3 \Lambda \,.
\eeq

\subsection{The linear problems and quantum Painlev\'e-Calogero
correspondence}

In this subsection we give a general view on what we are
going to do in sections 3--8 for the particular Painlev\'e equations.

In the original form, the Painlev\'e equations can be written as
\beq\label{lp1}
\p_T^2 y= R(T, y, \p_T y),
\eeq
where $R$ is a rational function of the independent variable
$T$, the dependent variable $y$ and its $T$-derivative.
The Painlev\'e-Calogero correspondence means
the existence of a change of variables from $y,T$ to $x,t$
of the form $y=y(x,t)$, $T=T(t)$
such that eq. (\ref{lp1}) in the new variables acquires the form
\beq\label{lp2}
\ddot x = -\p_x V(x,t)
\eeq
which is the Newton equation for motion of a point-like particle
on the line in a time-dependent potential $V(x,t)$.
In order to indicate the dependence on the parameters
$\alpha , \beta , \gamma , \delta$ which may enter the
Painlev\'e equations, we will write
$V(x,t)=V^{(\alpha , \beta , \gamma , \delta )}(x,t)$.
As it was already said in the Introduction,
we call (\ref{lp2}) {\it the Calogero form} of the Painlev\'e equation.
Hereafter, the dot means the $t$-derivative. It should be noted
that ${\rm P}_{\rm I}$ and ${\rm P}_{\rm II}$ equations are
already of the Calogero form, so no change
of the variables is necessary, for
${\rm P}_{\rm III}$ -- ${\rm P}_{\rm V}$ equations the
transformation $y \to x$ bringing the equations to the
Calogero form does not depend on $t$, and only for
${\rm P}_{\rm VI}$ this transformation is actually $t$-dependent.

The linear problems of the necessary form described in section
2.1 have been known for lower members of the Painlev\'e family
but not for higher ones (especially for ${\rm P}_{\rm V}$
and ${\rm P}_{\rm VI}$). Therefore, we should start from
a known version of the linear problems and then transform it
to the desired form. A convenient starting point is the pair
of compatible linear problems
\beq\label{P6JM1aa}
\left \{ \begin{array}{l}
\p_{X}\mathsf{\Psi} ={\sf U}(X,T)\mathsf{\Psi}
\\ \\
\p_T \mathsf{\Psi} ={\sf V}(X,T)\mathsf{\Psi}
\end{array}
\right.
\eeq
for a two-component vector function $\mathsf{\Psi}$,
where the matrices ${\sf U}(X,T)$, ${\sf V}(X,T)$ are rational
functions of the spectral parameter $X$ given in \cite{JM81-2}
for all the six Painlev\'e equations. The transformation
from this pair of matrices to the pair of matrices
${\bf U}(x,t)$, ${\bf V}(x,t)$
with the desired properties will be done in two steps:
$$
\{ {\sf U}(X,T), {\sf V}(X,T)\}\stackrel{{\cal R}}{\longrightarrow}
\{ {\bf \tilde U}(x,t), {\bf \tilde V}(x,t)\}
\stackrel{{\cal G}}{\longrightarrow}
\{ {\bf U}(x,t), {\bf V}(x,t)\}.
$$

The transformation ${\cal R}$ is a re-parametrization of the
time and spectral parameter corresponding to the change of variables
that prepares the Calogero form of the Painlev\'e equation
from the original one.
Here are some general relations for a change of variables
from $X,T$ to $x,t$ of the form $X=X(x,t)$, $T=T(t)$.
Clearly, such a change of variables
implies the following relations for the partial derivatives:
$$
\p_x = \frac{\p X}{\p x} \, \p_X\,, \quad \quad
\p_t = \frac{\p X}{\p t}\, \p_X +\frac{\p T}{\p t}\, \p_T.
$$
This means that the linear problems (\ref{P6JM1})
are transformed as follows:
\beq\label{6p3}
\left \{\begin{array}{l}
\displaystyle{\p_x \mathsf{\Psi} =\frac{\p X}{\p x} \,
{\sf U} \mathsf{\Psi}}
\\ \\
\displaystyle{\p_t \mathsf{\Psi} =\left (\frac{\p T}{\p t} \, {\sf V}
+\frac{\p X}{\p t}\, {\sf U}\right ) \mathsf{\Psi}.}
\end{array}\right.
\eeq
Therefore, the $U$--$V$ pair in the variables $x,t$ is
\beq\label{6p311}
\begin{array}{lll}
{\bf \tilde U} (x,t)& =&\displaystyle{\frac{\p X}{\p x} \,
{\sf U}(X(x,t), T(t))}
\\ &&\\
{\bf \tilde V}(x,t) &=& \displaystyle{\frac{\p T}{\p t} \,
{\sf V}(X(x,t), T(t))
+\frac{\p X}{\p t}\, {\sf U}(X(x,t), T(t))},
\end{array}
\eeq
where the entries of the matrices ${\sf U}$, ${\sf V}$
in the right hand side should be expressed through the
new variables $x,t$.
Note that
we deliberately use the same
letter $x$ as in the equation of the classical motion (\ref{lp2})
to stress the fact that it is this variable ($x$-coordinate of
a particle on the line) which is going to be
``quantized'' in the ``quantum''
version of the Painlev\'e-Calogero correspondence in the
sense that the momentum
$p=\dot x$ is going to be replaced by the operator $\p_x$.
This analogy is justified by the final formulas.

The zero curvature
condition for the pair of matrices
${\bf \tilde U}(x,t)$, ${\bf \tilde V}(x,t)$
is equivalent to the Painlev\'e equation
in the Calogero form
\beq\label{lp3}
\ddot u = -\p_u V(u,t)
\eeq
for the function
$u=u(t)$ defined as a (simple) zero of the
right upper element $\tilde b(x,t)={\bf \tilde U}_{12}(x,t)$ of the matrix
${\bf \tilde U}$: $\tilde b(u,t)=0$. (To avoid a misunderstanding,
we should stress that the time dependence of the
function $u(t)$ is defined
{\it not} by this equation but by the Painlev\'e equation.)

In general, the so obtained matrices
${\bf \tilde U}(x,t)$, ${\bf \tilde V}(x,t)$ do not obey
the condition (\ref{Bb}).
The transformation ${\cal G}$ is a diagonal gauge transformation
of the form (\ref{tilde1}) with a specially adjusted function
$\omega (x,t)$ such that the condition (\ref{Bb})
for the gauge-transformed matrices is satisfied.
Here an important remark is in order.
Given any two $2\times 2$ matrix functions
${\bf \tilde U}(x,t)$, ${\bf \tilde V}(x,t)$, one can always
find a scalar function $\omega (x,t)$ such that the upper right
entries of the gauge-transformed matrices,
$b=\tilde b \omega^2={\bf \tilde U}_{12}\omega^2$,
$B=\tilde B \omega^2={\bf \tilde V}_{12}\omega^2$,
are related by the equation $b_x =2B$.
Indeed, such a function $\omega$ can be found as a solution
to the differential equation
$$
\p_x \log \omega = \frac{\tilde B}{\tilde b} -
\frac{1}{2}\, \p_x \log \tilde b\,.
$$
A non-trivial additional constraint on the function $\omega$
is that it should factorize into a product of two functions
such that one of them depends on $x,t$ but {\it does not contain
the dependent variable $u$} and another one depends
on $t$ only (through both dependent and independent variables).
In fact this is a necessary condition for the
perfect classical-quantum correspondence.
Otherwise the spectral parameter and the dependent variable
have no chance to separate in the potential of the
non-stationary Schr\"odinger equation.
In fact the example of ${\rm P}_{\rm VI}$
shows that the two transformations, ${\cal R}$ and ${\cal G}$,
should be found simultaneously from the condition that the function
$\omega$ be of the special form in which the dependent variable
separates from the spectral parameter.

The resulting pair of matrices ${\bf U}(x,t)$, ${\bf V}(x,t)$
is the one that was discussed in section 2.1. The zero curvature
condition for these matrices is equivalent to the Painlev\'e equation
(\ref{lp3}) for the function
$u=u(t)$ which can be equivalently defined as a (simple) zero of the
right upper element $b(x,t)={\bf U}_{12}(x,t)$ of the matrix
${\bf U}$: $b(u,t)=0$. One can also check that the value of
the diagonal element, ${\bf U}_{11}(x,t)$, at $x=u$ is a canonically
conjugate variable to $u$, in accordance with the general constructions
of \cite{VN84,Sklyanin}. (A more detailed discussion of this point
will be given elsewhere.)

Further, we are going to reduce the system of linear problems
(\ref{d1}) to the pair of scalar
Schr\"odinger-like equations (\ref{scalareqs}) according to
the procedure outlined in sections 2.2 and 2.3.
The result merits attention and further understanding
from ``first principles''. The explicit calculations in each case
show that for any Painlev\'e equation (with possible
parameters $\alpha , \beta , \gamma , \delta $) the following
holds true:
\begin{itemize}
\item
The variables $x,u$ separate in the
non-stationary Schrodinger equation
meaning that
\beq\label{lp4}
U(x,t)=V^{(\tilde \alpha ,
\tilde \beta , \tilde \gamma , \tilde \delta )}(x,t)-
H^{(\alpha , \beta , \gamma , \delta )} (\dot u, u),
\eeq
where the potential $V^{(\tilde \alpha ,
\tilde \beta , \tilde \gamma , \tilde \delta )}(x,t)$
is of the same form as the one
for the classical equation (\ref{lp2}) (or (\ref{lp3}))
with possibly modified parameters and the $x$-independent
term, $H^{(\alpha , \beta , \gamma , \delta )} (\dot u, u)$,
is the classical Hamiltonian
$$
H(\dot u, u)=H^{(\alpha , \beta , \gamma , \delta )} (\dot u, u)=
\frac{1}{2}\, \dot u^2 + V^{(\alpha , \beta , \gamma , \delta )}(u,t)
$$
for the Painlev\'e equation in the Calogero form;
\item For ${\rm P}_{\rm I}$--${\rm P}_{\rm III}$
the parameters in the quantum Hamiltonian
are the same as in the classical one while for
${\rm P}_{\rm IV}$--${\rm P}_{\rm VI}$ some or all
parameters should be shifted:
$(\tilde \alpha , \tilde \beta  ) =(\alpha ,
\beta +\frac{1}{2})$ for ${\rm P}_{\rm IV}$,
$(\tilde \alpha ,
\tilde \beta , \tilde \gamma , \tilde \delta )=
(\alpha -\frac{1}{8}, \beta +\frac{1}{8}, \gamma , \delta )$
for ${\rm P}_{\rm V}$ and
$(\tilde \alpha ,
\tilde \beta , \tilde \gamma , \tilde \delta )=
(\alpha -\frac{1}{8}, \beta +\frac{1}{8},
\gamma -\frac{1}{8}, \delta +\frac{1}{8})$ for ${\rm P}_{\rm VI}$.
\end{itemize}
This means that the function
\beq\label{lp5}
\Psi (x,t)=e^{\int^t H(\dot u , u)dt'}\psi (x,t)
\eeq
is a common solution to the linear differential equations
\beq\label{scalareqs1}
\left \{
\begin{array}{l}\displaystyle{
\left (\frac{1}{2}\, \p_{x}^{2} -\frac{1}{2}\,
(\p_x \log b) \, \p_x \,  + W (x,t)\right )\Psi =0}
\\ \\
\displaystyle{
\p_t \Psi =\left ( \frac{1}{2}\, \p_x^2 +V(x,t)\right )\Psi .}
\end{array}
\right.
\eeq
The second one is
the non-stationary Schr\"odinger equation
$\p_t \Psi  =H(\p_x , x)\Psi$ whose Hamiltonian is the
natural quantization of the classical Hamiltonian of the
Painlev\'e equation, possibly with modified parameters
(such a modification, if any, can be regarded as a ``quantum
correction''). This is what we call the quantum
Painlev\'e-Calogero correspondence or the classical-quantum
correspondence for the Painlev\'e equations.

\section{Painlev\'e I}

\subsection{The equation}

The ${\rm P}_{\rm I}$ equation
\beq\label{P1}
4\ddot x=6x^2 +t
\eeq
is already of the Calogero form from the very beginning,
so no change of variables is necessary in this case.
It can be written in the standard Hamiltonian form as
$$
\dot x =\frac{\p H_{\rm I}}{\p p}\,, \quad
\quad \dot p =-\frac{\p H_{\rm I}}{\p x}
$$
with the classical time-dependent Hamiltonian
\beq\label{P1a}
H_{\rm I}= H_{\rm I}(p, x)=\frac{p^2}{2} -\frac{x^3}{2}
-\frac{tx}{4}\,.
\eeq
One may introduce the potential
\beq\label{P1b}
V_{\rm I}(x)=-\frac{x^3}{2}
-\frac{tx}{4}\,,
\eeq
then the ${\rm P}_{\rm I}$ equation takes the form
$\ddot x =-\p_x V_{\rm I}(x)$ which is the Newton equation
for a point-like particle on the line in the time-dependent
potential.
Note that the partial and full time derivatives of the
Hamiltonian coincide:
\beq\label{pf}
\frac{\p H_{\rm I}}{\p t}=\frac{d H_{\rm I}}{d t}=
-\, \frac{x(t)}{4}
\eeq
(the first equality is of course the general property
of Hamiltonians for non-conservative systems while the second one
is specific for the ${\rm P}_{\rm I}$ equation).

\subsection{Linearization and
classical-quantum correspondence for ${\rm P}_{\rm I}$}

In the case of ${\rm P}_{\rm I}$ the general construction
outlined in section 2 is especially simple and transparent
because it does not need neither the change of variables nor
the gauge transformation.
The ${\rm P}_{\rm I}$ equation
$4\ddot u=6u^2+t$
is known to be the compatibility condition for the
linear problems (\ref{d1})
with the matrices
\beq\label{matricesU}
{\bf U}(x,t)= \left (
\begin{array}{cc}
\dot u & x-u\\ &\\
x^2 \! +\! xu \! +\! u^2 \! +\! \frac{1}{2}\,t &  -\dot u
\end{array}\right ),\quad \quad
{\bf V} (x,t) =\left ( \begin{array}{cc}
0& \frac{1}{2}\\ &\\
\frac{1}{2}\, x +u & 0
\end{array}\right )
\eeq
which are already of the form implied in section 2.1.
Note that $u$ is the
simple zero of the right upper element of the matrix
${\bf U}(x,t)$: $b(u)=0$.

Another meaning of the ${\rm P}_{\rm I}$ equation
(which we will not discuss here) is the condition
that the monodromy data of the first linear problem in (\ref{d1})
be independent of the parameter $t$.

The spectral parameter is denoted by $x$. We deliberately use the same
letter $x$ as in the equation of the classical motion (\ref{P1})
to stress the fact that it is this variable ($x$-coordinate of
a particle on the line) which is going to be
``quantized'' in the ``quantum''
version of the Painlev\'e-Calogero correspondence in the
sense that the momentum
$p=\dot x$ is going to be replaced by the operator $\p_x$.
The notation with the same idea in mind will be used below for other
Painlev\'e equations.

It remains to apply the general formulas of section 2.
Consider equation (\ref{P1eea}).
In the case of ${\rm P}_{\rm I}$
$$
ad-bc= -x^3 -\frac{tx}{2}-\dot u^{2}+u^3 +\frac{tu}{2}\,,
\quad \quad
a_x = A =0.
$$
so the calculation of the potential $U(x,t)$ is
very simple.
As a result, we obtain the non-stationary
Schr\"odinger equation (in imaginary time)
\beq\label{P1f}
\p_t \psi =\left (\frac{1}{2}\, \p_{x}^{2}-\frac{x^3}{2}-
\frac{tx}{4}-H_{\rm I}(\dot u, u)\right )\psi \,,
\eeq
where
$H_{\rm I}(\dot u, u)$ is given by (\ref{P1a}).
We can write it in the form
\beq\label{P1g}
\p_t \psi =\Bigl (H_{\rm I}(\p_x , x)-H_{\rm I}(\dot u, u)
\Bigr )\psi
\eeq
where
$$
H_{\rm I}(\p_x , x)=\frac{1}{2}\,\p_{x}^{2}-\frac{x^3}{2}-
\frac{tx}{4}
$$
is the {\it quantum} Hamiltonian operator obtained as a literal
quantization of the classical Hamiltonian (\ref{P1a}).
The function
\beq\label{P1h}
\Psi (x,t)=e^{\int^t H_{\rm I}(\dot u , u)dt'}\psi (x,t)
\eeq
thus obeys the non-stationary Schr\"odinger equation
\beq\label{P1i}
\p_t \Psi = H_{\rm I}(\p_x , x)\Psi
=\left ( \frac{1}{2}\, \p_x^2 +V_{\rm I}(x,t)\right )\Psi
\eeq
without a free $t$-dependent term.

To conclude, we have two equivalent representations
of the ${\rm P}_{\rm I}$ equation. One is a classical motion in the
time-dependent cubic potential
with Hamiltonian
(\ref{P1a}). The coordinate of the particle as a function
of time obeys the ${\rm P}_{\rm I}$ equation. Another
representation is a time-dependent
quantum mechanical particle in the same time-dependent potential.
The non-stationary Schr\"odinger equation for this quantum
system in the coordinate representation simultaneously serves
as the linear problem
for time evolution associated with
the Painlev\'e equation\footnote{As we learned from B.Suleimanov
after completetion of this work, this fact was pointed out in
\cite{Suleimanov94,Suleimanov08}, see also \cite{DNovikov09}.}.

\section{Painlev\'e II}

\subsection{The equation}

The ${\rm P}_{\rm II}$ equation
\beq\label{P2}
\ddot x=2x^3 +tx-\alpha \,,
\eeq
where $\alpha$ is an arbitrary parameter,
is already of the Calogero form from the very beginning,
so no change of variables is necessary in this case.
It can be written in the standard Hamiltonian form as
$$
\dot x =\frac{\p H_{\rm II}}{\p p}\,, \quad
\quad \dot p =-\frac{\p H_{\rm II}}{\p x}
$$
with the classical time-dependent Hamiltonian
\beq\label{P2aa}
\begin{array}{lll}
H_{\rm II}= H_{\rm II}(p, x)&=&
\displaystyle{\frac{p^2}{2} -\frac{1}{2}
\left (x^2+\frac{t}{2}\right )^2
+\alpha x}
\\ &&\\
&=&\displaystyle{\frac{p^2}{2}-\frac{x^4}{2}-\frac{tx^2}{2}-
\frac{t^2}{8}+\alpha x}.
\end{array}
\eeq
One may introduce the potential
\beq\label{P2bb}
V_{\rm II}(x)=-\frac{1}{2}\left (x^2+\frac{t}{2}\right )^2
+\alpha x,
\eeq
then the ${\rm P}_{\rm II}$ equation takes the Newton form
$\ddot x =-\p_x V_{\rm II}(x)$.
Note that the partial and full time derivatives of the
Hamiltonian coincide:
\beq\label{pfa}
\frac{\p H_{\rm II}}{\p t}=\frac{d H_{\rm II}}{d t}=
-\, \frac{x^2(t)}{2}-\, \frac{t}{4}
\eeq
(again, the first equality is a general property
of Hamiltonians for non-conservative systems while the second one
is specific for the ${\rm P}_{\rm II}$ equation).

\subsection{Linearization and
classical-quantum correspondence for ${\rm P}_{\rm II}$}

The linear problems and their compatibility
condition
for the ${\rm P}_{\rm II}$ equation
$$
\ddot u=2u^3 +tu-\alpha
$$
are given by (\ref{d1}), (\ref{P1c}) with the matrices
\beq\label{matricesU2}
{\bf U}= \left (
\begin{array}{cc}
x^2+\dot u -u^2 & x-u\\ &\\
(x+u)(2u^2 \! -\! 2\dot u \! +\! t)\! -\!
2\alpha \! - \! 1 &  -x^2\! -\! \dot u \!  +\! u^2
\end{array}\right ),\quad
{\bf V}  =\left (
\begin{array}{cc}
\frac{x+u}{2}& \frac{1}{2}\\ &\\
u^2 \! - \! \dot u \! +\! \frac{t}{2} & -\,
\frac{x+u}{2}\end{array}\right ).
\eeq
They are of the form implied in section 2.1.
Note that $u$ is the
simple zero of the right upper element of the matrix
${\bf U}(x,t)$: $b(u)=0$.

The spectral parameter is again deliberately denoted
by the same letter $x$ as in the equation of
classical motion (\ref{P2})
to stress the fact that it is this variable ($x$-coordinate of
a particle on the line) which is going to be
``quantized'' in the ``quantum''
version of the Painlev\'e-Calogero correspondence in the
sense that the momentum
$p=\dot x$ is going to be replaced by the operator $\p_x$.

Another meaning of the ${\rm P}_{\rm II}$ equation
(which we will not discuss here) is the condition
that the monodromy data of the first linear problem in (\ref{d1})
be independent of the parameter $t$.

It remains to calculate the potential $U(x,t)$ of the
non-stationary Schr\"odinger equation.
In the case of ${\rm P}_{\rm II}$
$$
\begin{array}{rcl}
ad-bc&=& \displaystyle{-x^4 -tx^2 +(2\alpha +1)x
-\dot u^{2}+u^4 +tu^2 -(2\alpha +1)u\,, }
\\ &&\\
a_x -b_x A/B &=&x-u.
\end{array}
$$
As a result, we obtain the non-stationary
Schr\"odinger equation (in imaginary time)
\beq\label{P2h}
\p_t \psi =\left (\frac{1}{2}\, \p_{x}^{2}-\frac{x^4}{2}-
\frac{tx^2}{2}+\alpha x -\frac{t^2}{8}-H_{\rm II}(\dot u, u)\right )\psi
\eeq
or
\beq\label{P2g}
\p_t \psi =\Bigl (H_{\rm II}(\p_x , x)-H_{\rm II}(\dot u, u)
\Bigr )\psi ,
\eeq
where
$$
H_{\rm II}(\p_x , x)=\frac{1}{2}\, \p_{x}^{2}-\frac{x^4}{2}-
\frac{tx^2}{2}+\alpha x -\frac{t^2}{8}
$$
is the {\it quantum} Hamiltonian operator obtained as a literal
quantization of the classical Hamiltonian (\ref{P2aa}).
The function
\beq\label{P2i}
\Psi (x,t)=e^{\int^t H_{\rm II}(\dot u , u)dt'}\psi (x,t)
\eeq
thus obeys the non-stationary Schr\"odinger equation
\beq\label{P2j}
\p_t \Psi = H_{\rm II}(\p_x , x)\Psi
=\left ( \frac{1}{2}\, \p_x^2 +V_{\rm II}(x,t)\right )\Psi
\eeq
without a free $t$-dependent term.

To conclude, we have two equivalent representations
of the ${\rm P}_{\rm II}$ equation. One is a classical motion in the
time-dependent polynomial potential
with Hamiltonian
(\ref{P2aa}). The coordinate of the particle as a function
of time obeys the ${\rm P}_{\rm II}$ equation. Another
representation is a
quantum mechanical particle in the same time-dependent potential.
The non-stationary Schr\"odinger equation for this quantum
system in the coordinate representation simultaneously serves
as the linear problem
for time evolution associated with the Painlev\'e equation.

\section{Painlev\'e IV}

\subsection{The equation}

The standard form of the Painlev\'e IV (${\rm P}_{\rm IV}$) equation is
\beq\label{P4}
\p^2_t y=\frac{(\p_t y)^2}{2y}+\frac{3}{2}\, y^3
+4ty^2 +2 (t^2 -\alpha ) y +\frac{\beta}{y}\,,
\eeq
where $\alpha$, $\beta$ are arbitrary parameters.
This is the first example where a change of variable
is necessary. The time variable $t$ remains the same
but the dependent variable should be changed as $y=x^2$.
This brings the equation
to the Newton form
\beq\label{P4a}
\ddot x=\frac{3}{4}\, x^5
+2tx^3 + (t^2 -\alpha ) x +\frac{\beta}{2x^3}
\eeq
which admits a Hamiltonian structure similar to the previous
examples:
$$
\dot x =\frac{\p H_{\rm IV}^{(\alpha , \beta )}}{\p p}\,, \quad
\quad \dot p =-\frac{\p H_{\rm IV}^{(\alpha , \beta )}}{\p x}
$$
with the classical time-dependent Hamiltonian
\beq\label{P4aa}
\begin{array}{lll}
H_{\rm IV}^{(\alpha , \beta )}= H_{\rm IV}^{(\alpha , \beta )}(p, x)&=&
\displaystyle{\frac{p^2}{2} -\frac{x^6}{8}
-\frac{tx^4}{2} -\frac{1}{2}\left (t^2 -\alpha \right )x^2
+\frac{\beta}{4x^2}\,.}
\end{array}
\eeq
One may introduce the potential
\beq\label{P4bb}
V_{\rm IV}(x)=V^{(\alpha , \beta )}_{\rm IV}(x) -\frac{x^6}{8}
-\frac{tx^4}{2} -\frac{1}{2}\left (t^2 -\alpha \right )x^2
+\frac{\beta}{4x^2},
\eeq
then the ${\rm P}_{\rm IV}$ equation in the Calogero form
(\ref{P4a}) reads
$\ddot x =-\p_x V_{\rm IV}(x)$.
Note that the partial and full time derivatives of the
Hamiltonian coincide:
\beq\label{pfb}
\frac{\p H^{(\alpha , \beta )}_{\rm IV}}{\p t}=
\frac{d H^{(\alpha , \beta )}_{\rm IV}}{d t}=
-\, \frac{x^4(t)}{2}- tx^2(t)
\eeq
(again, the first equality is a general property
of Hamiltonians for non-conservative systems while the second one
is specific for the ${\rm P}_{\rm IV}$ equation).

\subsection{Linearization and
classical-quantum correspondence for ${\rm P}_{\rm IV}$}

The system of linear problems associated with
the ${\rm P}_{\rm IV}$ equation
for the $u$-variable in the Calogero form,
\beq\label{P4au}
\ddot u=\frac{3}{4}\, u^5
+2tu^3 + (t^2 -\alpha ) u +\frac{\beta}{2u^3}\,,
\eeq
is a modified version of the one given in \cite{Joshi06}.
Their compatibility
condition
is of the same form (\ref{P1c}) with
\beq\label{matricesU4a}
{\bf U}= \left (
\begin{array}{cc}\displaystyle{
\frac{x^3}{2}\! +\! tx \! +\! \frac{Q+\frac{1}{2}}{x}}&x^2 -u^2
\\ & \\
\displaystyle{\frac{Q^2+\frac{\beta}{2}}{u^2x^2}\! -\! Q\! -\!
\alpha \! -\! 1} & \,\,\,\,\displaystyle{
-\frac{x^3}{2}\! -\! tx \! -\! \frac{Q+\frac{1}{2}}{x}}
\end{array}\right )
\eeq

\beq\label{matricesU4b}
{\bf V} =\left (
\begin{array}{cc}
\displaystyle{
\frac{x^2+u^2}{2} + t}&\,\, x
\\ & \\
\displaystyle{
-\, \frac{Q+\alpha +1}{x}}&\,\,\,\,\,\,\,\,
\displaystyle{
-\frac{x^2 +u^2}{2} - t}
\end{array}\right ),
\eeq
where
$$
Q=u\dot u -\frac{u^4}{2}-tu^2.
$$
Note that these matrices enjoy the property
$b_x =2B$ and, therefore, the non-stationary
Schr\"odinger equation of the form (\ref{P1eea}) is valid.
(Equivalently, we could start from a rational $U$--$V$ pair
given in \cite{JM81-2}) and transform it to the desired form
according to the strategy outlined in section 2.4 but
in this case the transformatons are simple enough and
do not require any special consideration.)
Note also that $u$ is one of the two
simple zeros of the right upper element of the matrix
${\bf U}(x,t)$: $b(u)=0$. The second zero at the point
$x=-u$ leads to the same results because the equation
(\ref{P4au}) is invariant under the transformation
$u\to -u$.

Again,
we deliberately denote the spectral parameter by the same
letter $x$ as in the equation of classical motion (\ref{P4a})
to stress the fact that it is this variable ($x$-coordinate of
a particle on the line) which is going to be
``quantized'' in the ``quantum''
version of the Painlev\'e-Calogero correspondence in the
sense that the momentum
$p=\dot x$ is going to be replaced by the operator $\p_x$.

Let us calculate the potential $U(x,t)$ in equation
(\ref{P1eea}). It consists of two parts:
one of them is one half of the determinant of the matrix ${\bf U}$
and another one is $-\frac{1}{2}\, a_x +A$.
For clarity, we present the results for these two parts separately
and then take the sum. The calculation of the determinant yields:
$$
\begin{array}{lll}
ad-bc& = & \displaystyle{\frac{x^6}{4}-tx^4 -
\Bigl (t^2 -\alpha -\frac{1}{2}\Bigr )x^2 +
\frac{\beta -\frac{1}{2}}{2x^2}}
\\ &&\\
&&\displaystyle{-\,\, \left ( \dot u^2 -\frac{u^6}{4}-tu^4 -
(t^2 -\alpha -1)u^2 +\frac{\beta}{2u^2}\right )-t-
\frac{Q}{x^2}\,.}
\end{array}
$$
We see that the variables $x$ and $u$ do not completely separate
in this expression because of the last term
(recall that $Q$ is not a constant but a dynamical variable).
Fortunately, this term cancels out after adding the second part
of the potential:
$$
-a_x +2A = -\frac{x^2}{2} +\frac{1}{2x^2}+u^2 +t+
\frac{Q}{x^2}\,.
$$
Combining the two parts together, we get:
\beq\label{detp4a}
\begin{array}{lll}
\displaystyle{\frac{1}{2}(ad-bc -a_x +2A)}&=&
\displaystyle{\frac{x^6}{8}-\frac{tx^4}{2} -
\frac{1}{2}\Bigl (t^2 -\alpha \Bigr )x^2 +
\frac{\beta +\frac{1}{2}}{4x^2}}
\\ &&\\
&&\displaystyle{\!\!\!\!\!\!\!\!
-\,\, \left ( \frac{\dot u^2}{2} -\frac{u^6}{8}-\frac{tu^4}{2} -
\frac{1}{2}(t^2 -\alpha )u^2 +\frac{\beta}{4u^2}\right ).}
\end{array}
\eeq
Therefore, equation (\ref{P1eea})
reads
\beq\label{P4g}
\p_t \psi =\Bigl (H_{\rm IV}^{(\alpha , \beta +\frac{1}{2})}(\p_x , x)-
H_{\rm IV}^{(\alpha , \beta )}(\dot u, u)
\Bigr )\psi ,
\eeq
where
$$
H_{\rm IV}^{(\alpha , \beta +\frac{1}{2})}(\p_x , x)=\frac{1}{2}\, \p_{x}^{2}-\frac{x^6}{8}-
\frac{tx^4}{2}-\frac{1}{2}(t^2-\alpha )x^2 +
\frac{\beta +\frac{1}{2}}{4x^2}\,.
$$
The function
\beq\label{P4i}
\Psi (x,t)=e^{\int^t H^{(\alpha , \beta )}_{\rm IV}
(\dot u , u)dt'}\psi (x,t)
\eeq
thus obeys the non-stationary Schr\"odinger equation
\beq\label{P4j}
\p_t \Psi = H_{\rm IV}^{(\alpha , \beta +\frac{1}{2})}(\p_x , x)\Psi
=\left ( \frac{1}{2}\, \p_x^2 +
V^{(\alpha , \beta +\frac{1}{2})}_{\rm IV}(x,t)\right )\Psi
\eeq
without a free $t$-dependent term.
Note the shift $\beta \to \beta +\frac{1}{2}$ which can be
thought of as a ``quantum correction''.

To conclude, we have two equivalent representations
of the ${\rm P}_{\rm IV}$ equation. One is a classical motion in the
time-dependent potential
with Hamiltonian
(\ref{P4aa}). The coordinate of the particle as a function
of time obeys the ${\rm P}_{\rm IV}$ equation. Another
representation is a
quantum mechanical particle in the time-dependent
potential of the same form,
with the modified coefficient in front of $1/x^2$.
The non-stationary Schr\"odinger equation for this quantum
system in the coordinate representation simultaneously serves
as the linear problem
for time evolution associated with the Painlev\'e equation.

\section{Painlev\'e III}

\subsection{The equation}

The standard form of the ${\rm P}_{\rm III}$ equation
for a function $y(T)$ is
\beq\label{P3}
\p^2_{T} y=\frac{(\p_{T} y)^2}{y}-\frac{\p_{T}y}{T}
+\frac{1}{T}(\alpha y^2 +\beta ) +\gamma y^3 +\frac{\delta}{y}\,,
\eeq
where $\alpha$, $\beta$, $\gamma$, $\delta$ are arbitrary parameters.
The change of the variables $T =e^t$, $y=e^{2x}$ brings this equation
to the Newton form
\beq\label{P3a}
2\ddot x=e^t(\alpha e^{2x} +\beta e^{-2x}) + e^{2t}(\gamma e^{4x}
+\delta e^{-4x}).
\eeq
Note that only two parameters of the four are really independent
because the other two can be eliminated by the shifts $x\to x_0$,
$t\to t_0$ with constant $x_0, t_0$. However, we will keep 3 parameters
in order to be able to consider some particular cases which are
not reachable otherwise. So, equation (\ref{P3a})
acquires the form
\beq\label{P3b}
\ddot x=2\nu^2 e^t\sinh (2x-2\varrho ) +4\mu^2 e^{2t}\sinh (4x),
\eeq
where $\nu , \mu , \varrho$ are the parameters. In principle,
one of them, say $\nu$ can be put equal to one by the shift
$t\to t -2\log \nu$ but this works only if $\nu \neq 0$.

Equation (\ref{P3b}) admits a Hamiltonian structure similar to the previous
examples:
$$
\dot x =\frac{\p H_{\rm III}}{\p p}\,, \quad
\quad \dot p =-\frac{\p H_{\rm III}}{\p x}
$$
with the classical time-dependent Hamiltonian
\beq\label{P3aa}
\begin{array}{lll}
H_{\rm III}= H_{\rm III}(p, x)&=&
\displaystyle{\frac{p^2}{2} -\nu^2 e^t \cosh (2x-2\varrho )
-\mu^2 e^{2t}\cosh (4x).}
\end{array}
\eeq
One may introduce the potential
\beq\label{P3bb}
V_{\rm III}(x)=-\nu^2 e^t \cosh (2x-2\varrho ) -\mu^2 e^{2t}
\cosh (4x),
\eeq
then the ${\rm P}_{\rm III}$ equation reads
$\ddot x =-\p_x V_{\rm III}(x)$.

The case $\mu =0$ is special.
In this case, one can put $\varrho =0$ without loss of generality, so
the ${\rm P}_{\rm III}$ equation acquires the form
\beq\label{P3bbb}
\ddot x=2\nu^2 e^t\sinh (2 x )
\eeq
with just one parameter $\nu$ (which in fact can be
eliminated by a shift of time) and with the classical Hamiltonian
\beq\label{P3aaa}
H_{\rm III}=
\frac{p^2}{2} -\nu^2 e^t \cosh (2x).
\eeq
This equation will be referred to as truncated
${\rm P}_{\rm III}$ equation.

\subsection{The $U$--$V$ pairs for ${\rm P}_{\rm III}$}

\subsubsection{The case of the truncated ${\rm P}_{\rm III}$ equation}

The truncated
${\rm P}_{\rm III}$ equation
\beq\label{tr111}
\ddot u =2\nu^2 e^t\sinh (2u )
\eeq
is the compatibility condition for linear problems
with matrices ${\bf U},{\bf V}$ of rather simple form.
Indeed, it is easy to check that the
zero curvature condition (\ref{P1c}) with
\beq\label{matricesU3a}
{\bf U}(x,t) = \left (
\begin{array}{cc}
\dot u &2\nu e^{t/2}\sinh (x-u )
\\ & \\
2\nu e^{t/2}\sinh (x+u )  & -\dot u
\end{array}\right ),
\eeq

\beq\label{matricesU3b}
{\bf V }(x,t) =\left (
\begin{array}{cc}
0& \nu e^{t/2}\cosh (x-u )
\\ & \\
\nu e^{t/2} \cosh (x+u )& 0
\end{array}\right )
\eeq
yields equation (\ref{tr111})\footnote{The Lax pairs
for ${\rm P}_{\rm III}$ and ${\rm P}_{\rm V}$
were obtained by G.Aminov and
S.Arthamonov via trigonometric scaling limits from the one found in
\cite{Zotov04}.
However, the condition $b_x =2B$ for the Lax pairs obtained in this way
holds in the case of the truncuted ${\rm P}_{\rm III}$ equation only and
does not hold in general.}.
Moreover, these matrices
obviously satisfy the condition $b_x =2B$
and $u$ is the first order
zero of the element $b(x)$ (of course there are infinitely many
zeros in the complex $x$-plane at the points of the lattice $u+\pi i
\ZZ$ but all of them obey the same equation
(\ref{tr111})).

\subsubsection{The general case}

In the general case the $U$--$V$ pair for the ${\rm P}_{\rm III}$ equation
is more complicated. We take the linear problems for ${\rm P}_{\rm III}$
given in \cite{JM81-2} as a starting point,
passing to the exponential parametrization from the
very beginning and then transform them
to the ones appropriate for our purpose.

So, we start with the linear problems
\beq\label{gen1}
\p_x \mathbf{\tilde \Psi} =\left (
\begin{array}{cc}
e^{2x+t}-g_{11}e^{-2x+t}+\theta \! +\! \frac{1}{2} &
2v e^{x-\frac{t}{2}}-g_{12}e^{-x+\frac{t}{2}}
\\ &\\
2w e^{-x+\frac{t}{2}}-g_{21}e^{-3x+\frac{3t}{2}}&
\,\,\, -e^{2x+t}+g_{11}e^{-2x+t}-\theta \! -\! \frac{1}{2}
\end{array}\right )\mathbf{\tilde \Psi}
\eeq

\beq\label{gen2}
\p_t \mathbf{\tilde \Psi} =\frac{1}{2}\left (
\begin{array}{cc}
e^{2x+t}+g_{11}e^{-2x+t} - \frac{1}{2} &
\,\,\, 2v e^{x-\frac{t}{2}}+g_{12}e^{-x+\frac{t}{2}}
\\ &\\
2w e^{-x+\frac{t}{2}}+g_{21}e^{-3x+\frac{3t}{2}}&
\,\,\, -e^{2x+t}-g_{11}e^{-2x+t} + \frac{1}{2}
\end{array}\right )\mathbf{\tilde \Psi}
\eeq
where $v,w, g_{11}, g_{12}, g_{21}$ are yet unknown functions
of $t$ and
$\theta$ is a parameter.
The functions $g_{ik}$ are naturally thought of as
entries of a traceless matrix
\beq\label{Gless}
G=
\left (\begin{array}{cc}g_{11}&g_{12}
\\ g_{21}&-g_{11}\end{array}\right ).
\eeq
Note that the $x$-derivative of the right upper element of
the ${\bf \tilde U}$-matrix in (\ref{gen1}) is just equal to twice the
right upper element of
the ${\bf \tilde V}$-matrix in (\ref{gen2}).

As before, we deliberately denote the spectral parameter by the same
letter $x$ as in the equation of the classical motion (\ref{P3a})
to stress the fact that it is this variable ($x$-coordinate of
a particle on the line) which is going to be
``quantized'' in the ``quantum''
version of the Painlev\'e-Calogero correspondence in the
sense that the momentum
$p=\dot x$ is going to be replaced by the operator $\p_x$.

The compatibility of the
linear problems (\ref{gen1}),(\ref{gen2}) implies the
following system of differential equations:
\beq\label{gen3}
\left \{ \begin{array}{l}
\dot g_{11}=2(vg_{21}-wg_{12})
\\ \\
\dot g_{12}=\theta g_{12}-4vg_{11}
\\ \\
\dot g_{21}=-\theta g_{21}+4wg_{11}
\\ \\
\dot v \,\, = -\theta v -g_{12}e^{2t}
\\ \\
\dot w \,\, = \theta w +g_{21}e^{2t}.
\end{array}
\right.
\eeq
Combining these equations, one easily finds two integrals:
\beq\label{gen4}
\!\!\!\!\! \chi := g_{11}^{2}+g_{12}g_{21}
\eeq
\beq\label{gen5}
\,\,\,\,\,\,\,\,\,\,\lambda := vg_{21}+wg_{12}+\theta g_{11}\,,
\eeq
where $\chi$ and $\lambda$ are integration constants.
Note that the first integral is just determinant of the matrix
$G$ (\ref{Gless}) with opposite sign.

Using (\ref{gen4}), (\ref{gen5}), one can exclude $w$ and
$g_{21}$,
$$
w=\frac{\lambda -\theta g_{11}-vg_{21}}{g_{12}}\,,
\quad \quad g_{21}=\frac{\chi -g_{11}^{2}}{g_{12}},
$$
and reduce the system (\ref{gen3}) to a simpler one:
\beq\label{gen6}
\left \{
\begin{array}{l}
\dot g_{11}=4vg_{12}^{-1}(\chi - g_{11}^{2})+2\theta g_{11}
-2\lambda
\\ \\
\dot g_{12}=\theta g_{12}-4vg_{11}
\\ \\
\dot v \,\, = -\theta v -g_{12}e^{2t}.
\end{array}\right.
\eeq
Further, these equations imply the following system
for the functions $f=v/g_{12}$, $g=g_{11}$:
\beq\label{gen7}
\left \{ \begin{array}{l}
\dot f = 4gf^2 -2\theta f -e^{2t}
\\ \\
\dot g = -4fg^2 +2\theta g +4\chi f -2\lambda \,.
\end{array} \right.
\eeq
Now, substituting
$$
g=\frac{\dot f +2\theta f +e^{2t}}{4f^2}
$$
from the first equation into the second one,
we obtain a closed equation for
$f$,
\beq\label{gen8}
\ddot f = \frac{\dot f^{2}}{f}+16 \chi f^3 -8\lambda f^2
-2(\theta \! +\! 1)e^{2t}-\frac{e^{4t}}{f}
\eeq
which is equivalent to the ${\rm P}_{\rm III}$ equation
(\ref{P3}) and can be brought to the original form by the change
of variable $T= e^t$. The change of the dependent
variable $f=e^{-2u +t}$ yields the equation
\beq\label{gen9}
\ddot u =e^t \Bigl ((\theta \! +\! 1) e^{2u}
+4\lambda e^{-2u}\Bigr ) +
\frac{1}{2}\, e^{2t}\Bigl (e^{4u} -16\chi
e^{-4u}\Bigr )
\eeq
which has the form (\ref{P3b}) with $\mu =1/2$
under the identification of parameters
$$
\theta +1 = \nu^2 e^{-2\varrho}, \quad
4\lambda = -\nu^2 e^{2\varrho}, \quad
\chi =\frac{1}{16}\,.
$$

\subsection{Classical-quantum correspondence for ${\rm P}_{\rm III}$}

\subsubsection{The case of truncated ${\rm P}_{\rm III}$ equation}

Let us start with the simplest case $\mu =0$ and use the
$U$--$V$ pair (\ref{matricesU3a}), (\ref{matricesU3b}).
A simple calculation shows that in this case
the linear equation for $\psi$ (\ref{P1e})
becomes the ``non-stationary Mathieu equation''
\beq\label{P3g}
\p_t \psi =\Bigl (H_{\rm III}(\p_x , x)-
H_{\rm III}(\dot u , u)
\Bigr )\psi \,,
\eeq
where
$$
H_{\rm III}(\p_x , x)=\frac{1}{2}\, \p_{x}^{2}
-\nu^2 e^t \cosh (2x),
$$
i.e., we again observe a perfect classical-quantum correspondence.
Note that in this case $a_x =A=0$, so the potential is given
solely by determinant of the matrix ${\bf U}$:
$$
\begin{array}{lll}
\displaystyle{\frac{1}{2}(ad-bc)} & = &
\displaystyle{-\frac{\dot u ^2}{2}-2\nu ^2e^t
\sinh (x+u )\sinh (x-u )}
\\ && \\
& =&\displaystyle{-\nu ^2 e^t \cosh (2x)-
\left (\frac{\dot u ^2}{2}-
\nu ^2 e^t \cosh (2u )\right ).}
\end{array}
$$
We remark that the non-stationary Mathieu equation in connection with
the ${\rm P}_{\rm III}$ equation was mentioned in \cite{BM06}.

\subsubsection{The general case}

In order to achieve a precise classical-quantum correspondence
in the general case of the ${\rm P}_{\rm III}$ equation
with arbitrary parameters, one should modify the system of linear
problems given above by a diagonal
$x$-independent (but $t$-dependent) gauge transformation
of the form (\ref{tilde1}) with
\beq\label{gen10}
\omega = g_{12}^{-\frac{1}{2}} (2fe^{-t})^{-\frac{1}{4}},
\eeq
where $f=v/g_{12}$ as before.

Another small modification which is necessary to
achieve perfect classical-quantum correspondence
is the shift of the spectral parameter
$x\to x-\frac{1}{2}\log 2$. Then the linear problems
(\ref{gen1}), (\ref{gen2}) acquire the form (\ref{d1})
with
\beq\label{gen11}
{\bf U}=\left (
\begin{array}{cc}
\frac{1}{2}\, e^{2x+t}-2g_{11}e^{-2x+t}+\theta \! +\! \frac{1}{2} &
\,\,\,\,\, f^{\frac{1}{2}}\, e^{x}-
f^{-\frac{1}{2}}\,e^{-x+t}
\\ &\\
4(vg_{12})^{\frac{1}{2}}\Bigl (w e^{-x}-
g_{21}e^{-3x+t}\Bigr )&
\,\,\, -\frac{1}{2}\, e^{2x+t}+2g_{11}e^{-2x+t}-\theta \! -\! \frac{1}{2}
\end{array}\right )
\eeq

\beq\label{gen12}
{\bf V}=\left (
\begin{array}{cc}
\frac{1}{4}\, e^{2x+t}+g_{11}e^{-2x+t}+h &
\,\,\,\, \frac{1}{2}\Bigl (f^{\frac{1}{2}}\, e^{x}+
f^{-\frac{1}{2}}\, e^{-x+t}\Bigr )
\\ &\\
2(vg_{12})^{\frac{1}{2}}
\Bigl (w e^{-x}+g_{21}e^{-3x+t}\Bigr )&
\,\,\,\,\,\,\,\,\, -\frac{1}{4}\, e^{2x+t}-g_{11}e^{-2x+t} -h
\end{array}\right ),
\eeq
where
\beq\label{gen13}
h:= \p_t \log \Bigl (f^{-\frac{1}{4}}\, g_{12}^{-\frac{1}{2}}\Bigr )
=\frac{\dot f}{4f}+\frac{e^{2t}}{2f}+\frac{\theta}{2}\,.
\eeq
Recall also that
$$
f=e^{-2u +t},
$$
so the right upper element of the matrix ${\bf U}$ is
$b(x)=2e^{t/2} \sinh (x-u)$ and $u$ is its first order zero.

Now,
the calculation of the potential $U(x,t)$ in the non-stationary
Schr\"odinger equation (\ref{P1eea}) yields
\beq\label{gen14}
\begin{array}{lll}
\displaystyle{U(x,t)}&=&
\displaystyle{-\, \frac{e^{2t}}{8}\Bigl ( e^{4x}
+16 \chi e^{-4x}\Bigr )
-\, \frac{e^{t}}{2}\Bigl ( (\theta \! +\! 1)e^{2x}
-4\lambda  e^{-2x}\Bigr )}
\\ &&\\
&&\displaystyle{\!\!\!\!\!\!\!\!
-\,\, \frac{\dot u ^2}{2}
+\frac{e^{2t}}{8}\Bigl ( e^{4u }
+16 \chi e^{-4u}\Bigr )
+\frac{e^{t}}{2}\Bigl ( (\theta \! +\! 1)e^{2u}
-4\lambda  e^{-2u}\Bigr ),}
\end{array}
\eeq
so the Schr\"odinger equation acquires the desired form
\beq\label{gen15}
\p_t \psi =\Bigl (H_{\rm III}(\p_x , x)-
H_{\rm III}(\dot u, u )
\Bigr )\psi \,,
\eeq
where
$$
H_{\rm III}(\p_x , x)=\frac{1}{2}\,\p_{x}^{2}
-\, \frac{e^{2t}}{8}\Bigl ( e^{4x}
+16 \chi e^{-4x}\Bigr )
-\, \frac{e^{t}}{2}\Bigl ( (\theta \! +\! 1)e^{2x}
-4\lambda  e^{-2x}\Bigr ).
$$
The function
\beq\label{P3i}
\Psi (x,t)=e^{\int^t H_{\rm III}
(\dot u , u)dt'}\psi (x,t)
\eeq
thus obeys the non-stationary Schr\"odinger equation
\beq\label{P3j}
\p_t \Psi = H_{\rm III}(\p_x , x)\Psi
=\left ( \frac{1}{2}\, \p_x^2 +
V_{\rm III}(x,t)\right )\Psi
\eeq
with the same classical potential (\ref{P3bb}).

To conclude, we have two equivalent representations
of the ${\rm P}_{\rm III}$ equation. One is a classical motion in the
time-dependent potential
with Hamiltonian
(\ref{P3aa}). The coordinate of the particle as a function
of time obeys the ${\rm P}_{\rm III}$ equation. Another
representation is a
quantum mechanical particle in the same time-dependent
potential described by a non-stationary Schr\"odinger equation.
The latter simultaneously serves as the linear problem
for time evolution associated with the Painlev\'e equation.

\section{Painlev\'e V}

\subsection{The equation}

The standard form of the ${\rm P}_{\rm V}$ equation is
\beq\label{P5}
\p^2_{T} y=\left (\frac{1}{2y}+\frac{1}{y-1}\right )
(\p_{T} y)^2 -\frac{\p_{T} y}{T} +\frac{y(y-1)^2}{T^2}
\left (\alpha +\frac{\beta}{y^2}+\frac{\gamma T}{(y-1)^2}
+\frac{\delta T^2 (y+1)}{(y-1)^3}\right ),
\eeq
where $\alpha$, $\beta , \gamma , \delta$ are arbitrary parameters.
A re-scaling of the dependent variable allows one to fix one of these
parameters, so there are three essentially independent parameters.
The change of the time variable $T =e^{2t}$
allows one to eliminate the term $\p_{T}y/T$
in the right hand side,
so that the equation becomes
\beq\label{P5a}
\ddot y=\left (\frac{1}{2y}+\frac{1}{y-1}\right )
\dot y^2  +4(y-1)^2
\Bigl (\alpha y +\frac{\beta}{y}\Bigr  )
+4\gamma e^{2t}y
+\frac{4\delta e^{4t} y(y+1)}{y-1}\,.
\eeq
Further, the change of the dependent variable
\beq\label{P5aa}
y=\coth ^2 x
\eeq
brings the equation
to the form
\beq\label{P5aaa}
\ddot x =-\frac{2\alpha \cosh x}{\sinh^3 x}-
\frac{2\beta \sinh x}{\cosh^3 x}
-\gamma e^{2t}\sinh (2x) -\frac{1}{2}\, \delta e^{4t}\sinh (4x)
\eeq
which can be written as the Newton equation
\beq\label{P5aaaa}
\ddot x=-\p_x V_{\rm V}(x)
\eeq
with the time-dependent potential
\beq\label{P5aaab}
V_{\rm V}(x)=-\, \frac{\alpha}{\sinh^2 x}
-\, \frac{\beta}{\cosh^2 x}+\frac{\gamma e^{2t}}{2}
\cosh (2x)+\frac{\delta e^{4t}}{8}\cosh (4x).
\eeq
Again, we see that only three parameters among the four are really
independent because one of them can be put equal to 1 by a proper
shift of $t$.
This equation admits a Hamiltonian structure similar to the previous
cases:
$$
\dot x =\frac{\p H_{\rm V}}{\p p}\,, \quad
\quad \dot p =-\frac{\p H_{\rm V}}{\p x}
$$
with the classical time-dependent Hamiltonian
\beq\label{P5aab}
H_{\rm V}(p, x)=\frac{p^2}{2}+V_{\rm V}(x)
\eeq
To indicate the dependence on the parameters, we will
write $H_{\rm V}(p, x)=
H^{(\alpha , \beta , \gamma , \delta )}_{\rm V}(p, x)$.

\subsection{The zero curvature representation of
the ${\rm P}_{\rm V}$ equation}

The choice of the $U$--$V$ pair for the ${\rm P}_{\rm V}$ equation
suitable for our purpose is by no means obvious.
We start from a modified version of the
$U$--$V$ pair with rational dependence on the spectral
parameter suggested by M.Jimbo and T.Miwa \cite{JM81-2}
and then show how to transform it to the desired form.

\subsubsection{The modified Jimbo-Miwa
$U$--$V$ pair for ${\rm P}_{\rm V}$}

Let us consider the system of linear problems
\beq\label{JM1}
\left \{ \begin{array}{l}
\p_{X}\mathsf{\Psi} ={\sf U}(X,t)\mathsf{\Psi}
\\ \\
\p_t \mathsf{\Psi} ={\sf V}(X,t)\mathsf{\Psi}
\end{array}
\right.
\eeq
with the matrices
\beq\label{JM2}
{\sf U}=\left (
\begin{array}{cc}\displaystyle{\frac{e^{2t}}{2}+\frac{g}{X}
-\frac{g+\sigma}{X-1}} & \,\,\,\,\,\,
\displaystyle{\frac{v}{X}-\frac{w}{X-1}}
\\ &\\
\displaystyle{\frac{v_1}{X}-\frac{w_1}{X-1}}&\,\,\,\,\,\,
\displaystyle{-\frac{e^{2t}}{2}-\frac{g}{X}
+\frac{g+\sigma }{X-1}}
\end{array}\right )
\eeq

\beq\label{JM3}
\!\!\!\!\!\!\!\!\!\!{\sf V}=\left (
\begin{array}{cc}\displaystyle{Xe^{2t}} & \,\,\,\,\,\,
\displaystyle{2(v-w)}
\\ &\\
\displaystyle{2(v_1 -w_1)}&\,\,\,\,\,\,
\displaystyle{-Xe^{2t}}
\end{array}\right )
\eeq
and a column 2-component vector $\mathsf{\Psi}$.
Here $X$ is the spectral parameter, $v,w, v_1 , w_1, g$ are
some functions of $t$ to be constrained
by the zero curvature condition
$\p_X {\sf V}-\p_t {\sf U}+[{\sf V}, \,
{\sf U}]=0$ and $\sigma $ is an arbitrary constant.
The zero curvature condition yields the system of differential
equations
\beq\label{JM4}
\left \{
\begin{array}{l}
\dot g =2\, (vw_1-wv_1)
\\ \\
\dot v =-4(v-w)g
\\ \\
\dot w =-4(v-w)(g+\sigma )+2we^{2t}
\\ \\
\dot v_1 =4(v_1 -w_1 )g
\\ \\
\dot w_1 =4(v_1 -w_1 )(g+\sigma )-2w_1 e^{2t}.
\end{array}\right.
\eeq
Combining these equations, one easily finds two integrals:
\beq\label{JM5}
\begin{array}{l}
vv_1+g^2 =\zeta^2\,,
\\ \\
ww_1 +g(g+2\sigma )=\xi^2 +2\xi \sigma \,,
\end{array}
\eeq
where $\zeta$, $\xi$ are arbitrary constants
(the integration constants are expressed
in this particular way for later convenience).
These formulas allow one to substitute
$$
v_1 = \frac{(\zeta -g)(\zeta +g)}{v}\,, \quad \quad w_1 = \frac{(\xi
-g)(2\sigma + \xi +g)}{w}
$$
into the first equation of the system (\ref{JM4})
thus reducing it to the system of three equations
for three unknown functions.

Let us introduce the function
\beq\label{JM6}
y=\frac{v}{w}\,,
\eeq
then the first equation of the system (\ref{JM4}) becomes
\beq\label{JM7}
\dot g = 2\Bigl ( y^{-1}(g+\xi )(g-\xi )-y
(g+\xi +2\sigma )(g-\xi)\Bigr ).
\eeq
Writing
$\displaystyle{
\dot y = \frac{\dot v}{w}-\frac{v\dot w}{w^2}=
\frac{\dot v}{w}-y\frac{\dot w}{w}}\,,
$
we find from the second and third equations of the system (\ref{JM4}):
\beq\label{JM8}
\dot y = 4(y-1)^2 g +4\sigma y(y-1) -2y e^{2t}.
\eeq
Plugging
\beq\label{gyy}
g=\frac{\dot y +2ye^{2t}}{4(y-1)^2}\, -\, \frac{\sigma y}{y-1}
\eeq
expressed from this equation in terms of $y$ into
(\ref{JM7}), one obtains, after a relatively long calculation,
\beq\label{JM9}
\ddot y = \Bigl ( \frac{1}{2y}+\frac{1}{y\! -\! 1}\Bigr )\, \dot y^2
+8(y\! -\! 1)^2\Bigl ( (\xi +\sigma )^2 y -\frac{\zeta^2}{y}\Bigr )
 +4 (2\sigma \! -\! 1) e^{2t}y -\frac{2e^{4t}y(y+1)}{y-1}
\eeq
which is the ${\rm P}_{\rm V}$ equation in the form (\ref{P5a})
with
\beq\label{abgd}
\alpha = 2(\xi +\sigma )^2\,, \quad
\beta =-2\zeta^2\,, \quad
\gamma = 2\sigma -1\,, \quad
\delta = -\frac{1}{2}\,.
\eeq

\subsubsection{Hyperbolic parametrization}

The crucial step of the further construction is a parametrization
of the modified Jimbo-Miwa $U$--$V$ pair (\ref{JM2}), (\ref{JM3})
in terms of hyperbolic functions. This
pa\-ra\-met\-ri\-za\-tion
corresponds to the hyperbolic substitution (\ref{P5aa}) for the dependent
variable leading to the Calogero form of the
${\rm P}_{\rm V}$ equation but does not coincide with it.
The next step is a special diagonal
gauge transformation which recasts the matrices in the form
such that the condition $b_x=2B$ (\ref{Bb}) is satisfied.

The required hyperbolic parametrization is achieved by setting
\beq\label{p1}
X=\cosh ^2 x.
\eeq
Since this transformation does not depend on $t$,
the general formulas (\ref{6p3}) simplify.
Taking into account the rule $\p_x =2\cosh x \, \sinh x \, \p_X$
by which the derivative
$\p_X$ is transformed, we see that the first linear problem in
(\ref{JM1}) should be changed to $\p_x
\mathsf{\Psi} = 2\cosh x \, \sinh x \,
{\sf U}(X(x), t)\mathsf{\Psi}$, so the $U$--$V$ pair in the
hyperbolic parametrization acquires the form
\beq\label{p2}
{\bf \tilde U}(x,t)=\left ( \begin{array}{cc}
\tilde a &
2v \tanh x -2w \coth x
\\ &\\
\displaystyle{2v_1\tanh x -
2w_1 \coth x} &
-\, \tilde a
\end{array}
\right )
\eeq
with
$$
\tilde a=e^{2t}\sinh x \, \cosh x + 2g\tanh x -\Bigl (2g \! +\!
2\sigma \Bigr )\coth x
$$
and
\beq\label{p3}
{\bf \tilde V}(x,t)=\left ( \begin{array}{cc}
e^{2t}\cosh^2 x & \,\,\,\,\,\, 2(v-w)
\\ & \\
\displaystyle{2(v_1-w_1)}&
- \, e^{2t}\cosh^2 x
\end{array}\right ).
\eeq
Here the functions $v,w, v_1 , w_1, g$
are the same as in (\ref{JM2}), (\ref{JM3}).
Clearly, the zero curvature condition yields
equation (\ref{JM9}) with the same constants
$\zeta , \xi$ as in (\ref{JM5}). This $U$--$V$ pair
obeys the property $b_x =2B$.

We deliberately denote the
hyperbolic spectral parameter by the same
letter $x$ as in the equation of the classical motion (\ref{P5aaa})
to stress the fact that it is this variable ($x$-coordinate of
a particle on the line) which is going to be
``quantized'' in the ``quantum''
version of the Painlev\'e-Calogero correspondence in the
sense that the momentum
$p=\dot x$ is going to be replaced by the operator $\p_x$.

\subsection{Classical-quantum correspondence for ${\rm P}_{\rm V}$}

In order to achieve the precise classical-quantum
correspondence, one should apply a diagonal gauge transformation.
Namely, let us pass to the gauge equivalent
$U$--$V$ pair
$$
{\bf U}=\Omega ^{-1}{\bf \tilde U} \Omega -\Omega ^{-1}\p_x \Omega\,,
\quad \quad
{\bf V}=\Omega ^{-1}{\bf \tilde V} \Omega -\Omega ^{-1}\p_t \Omega
$$
with
$$
\Omega =\left (
\begin{array}{cc}\displaystyle{
\left (\frac{e^{-t} (v-w)}{\sinh x \, \cosh x}\right )^{\frac{1}{2}}} & 0
\\ & \\
0 & \displaystyle{\left (\frac{e^{-t} (v-w)}{\sinh x \,
\cosh x}\right )^{-\frac{1}{2}}}
\end{array}\right ).
$$
Explicitly, the $U$--$V$ pair (\ref{p2}), (\ref{p3})
transforms into
\beq\label{p2a}
{\bf U}(x,t)=\left ( \begin{array}{cc}
a &
\displaystyle{\frac{2e^t (v\sinh^2 x \! -\! w \cosh^2 x)}{v-w}}
\\ &\\
\displaystyle{2e^{-t}(v-w)\left (\frac{v_1}{\cosh^2 x} -
\frac{w_1}{\sinh^2 x}\right )} &
-\, a
\end{array}
\right )
\eeq
with
$$
a=e^{2t}\sinh x \, \cosh x +\Bigl (2g \! +\! \frac{1}{2}\Bigr )\tanh x
-\Bigl (2g \! +\! 2\sigma \! -\! \frac{1}{2}\Bigr )\coth x
$$
and
\beq\label{p3a}
{\bf V}(x,t) =\left (
\begin{array}{cc} e^{2t}\left(\cosh^2 x +\sinh^2 u\right)
\! -\! 2\sigma \! +\! \frac{1}{2} & \,\,\,\,\,\, e^t\sinh (2x)
\\ & \\
\displaystyle{\frac{4(v-w)(v_1-w_1)}{e^t\sinh (2x)}}&
-e^{2t}\left(\cosh^2 x+\sinh^2 u\right) \! +\! 2\sigma \! -\! \frac{1}{2}
\end{array}\right ).
\eeq
In principle, the auxiliary functions $g$, $v$, $w$,
$v_1$, $w_1$ can be excluded from the
hyperbolic $U$--$V$ pair (\ref{p2a}), (\ref{p3a}),
with the final result being written solely in terms of
$u$ and $\dot u$. However,
for the purpose of this paper we do not need this form
(it will be presented elsewhere).
The result of the previous subsections imply that setting
$$
y=\frac{v}{w}=\coth^2 u
$$
we find from the zero curvature condition for the
hyperbolic $U$--$V$ pair (\ref{p2a}), (\ref{p3a}) the
${\rm P}_{\rm V}$ equation in the Calogero-Inozemtsev-like form:
\beq\label{P5b}
\ddot u =-\frac{4(\xi +\sigma )^2\cosh u}{\sinh^3 u}+
\frac{4\zeta^2\sinh u}{\cosh^3 u}
-(2\sigma -1) e^{2t}\sinh (2u) +\frac{1}{4}\, e^{4t}\sinh (4u).
\eeq
Note that in this parametrization $b(x)=2e^t \sinh (x-u)\sinh (x+u)$.
For real $u$ this element has just two zeros
in the strip $|{\rm Im}\, x|<\pi$ at the points $\pm u$ and the
both obey the same equation (\ref{P5b}).

In order to check the classical-quantum correspondence,
one should calculate the potential $U(x,t)$ of the non-stationary
Schr\"odinger equation (\ref{P1eea}). The calculation
is straightforward and the result is
\beq\label{P5c}
\begin{array}{lll}
\displaystyle{U(x,t)}&=&
\displaystyle{\frac{4\zeta^2 -\frac{1}{4}}{2\cosh^2 x}-
\frac{4(\xi +\sigma )^2 -\frac{1}{4}}{2\sinh^2 x}}
\\ &&\\
&&\displaystyle{-\,\,\, \frac{e^{4t}}{16}\cosh (4x)+
\Bigl (\sigma -\frac{1}{2}\Bigr )e^{2t}\cosh (2x)}-\tilde H\,,
\end{array}
\eeq
where
\beq\label{P5d}
\tilde H = 2(v-w)(v_1
-w_1)-\frac{e^{4t}}{16}- e^{2t}\Bigl (2g +\frac{w}{v-w} \, + \!
\sigma \! +\! \frac{1}{2}\Bigr )+2\sigma ^2 .
\eeq
The $x$-dependent
part of the potential coincides with the potential (\ref{P5aaa}) up
to some shifts of the parameters $\alpha \to \alpha -\frac{1}{8}$,
$\beta \to \beta +\frac{1}{8}$ (see (\ref{abgd})). Let us find the
$x$-independent term $\tilde H$ and compare it with the classical
Hamiltonian for ${\rm P}_{\rm V}$. Using (\ref{JM5}), we get:
$$
2(v-w)(v_1 -w_1)=2(y-1)\Bigl (\frac{y-1}{y}\, g^2
+2\sigma g +\frac{\zeta^2}{y}-\xi^2 -2\xi \sigma \Bigr )
$$
and $g$ is given by (\ref{gyy}). Passing to the hyperbolic
parametrization, we have:
$$
g=-\frac{\dot u}{2}\, \sinh u \, \cosh u
+\frac{e^{2t}}{2}\, \sinh ^2u \, \cosh ^2u
-\sigma \cosh ^2 u
$$
and
$$
2(v-w)(v_1 -w_1)=\frac{\dot u ^2}{2}+
\frac{e^{4t}}{16}(\cosh (4u ) -1) -
\frac{e^{2t}}{2}\sinh (2u )\dot u +
\frac{2\zeta ^2}{\cosh ^2 u}
-\frac{2(\xi +\sigma )^2}{\sinh ^2 u}-2\sigma^2.
$$
Plugging all this into (\ref{P5d}), we get exactly the
classical Hamiltonian $H_{\rm V}^{(\alpha , \beta , \gamma ,
\delta )}(\dot u , u )$
with the parameters $\alpha , \beta ,\gamma , \delta$ given by
(\ref{abgd}):
\beq\label{P5e}
\begin{array}{lll}
\tilde H &= &\displaystyle{
\frac{\dot u ^2}{2}-\frac{2(\xi +\sigma )^2}{\sinh ^2 u}
+\frac{2\zeta ^2}{\cosh ^2 u}+
\frac{e^{2t}}{2}(2\sigma -1)\cosh (2u )-
\frac{e^{4t}}{16}\cosh (4u )}
\\ &&\\
&=&\displaystyle{H_{\rm V}^{(\alpha , \beta , \gamma ,
\delta )}(\dot u , u ).}
\end{array}
\eeq

Summing up, in the case of
${\rm P}_{\rm V}$ we have the non-stationary Schr\"odinger
equation
\beq\label{P5f}
\p_t \psi =\Bigl (H_{\rm V}^{(\alpha -\frac{1}{8}, \,
\beta +\frac{1}{8}, \, \gamma ,\,
\frac{1}{2} )}(\p_x , x)-
H_{\rm V}^{(\alpha , \beta , \gamma ,
\frac{1}{2} )}(\dot u , u )\Bigr )\psi .
\eeq
The function
\beq\label{P5i}
\Psi (x,t)=e^{\int^t H_{\rm V}^{(\alpha , \,
\beta , \, \gamma ,\,
\frac{1}{2} )}
(\dot u , u)dt'}\psi (x,t)
\eeq
thus obeys the non-stationary Schr\"odinger equation
\beq\label{P5j}
\p_t \Psi = H_{\rm V}^{(\alpha -\frac{1}{8}, \,
\beta +\frac{1}{8}, \, \gamma ,\,
\frac{1}{2} )}(\p_x , x)\Psi
=\left ( \frac{1}{2}\, \p_x^2 +
V_{\rm V}^{(\alpha -\frac{1}{8}, \,
\beta +\frac{1}{8}, \, \gamma ,\,
\frac{1}{2} )}(x,t)\right )\Psi .
\eeq
Note that the parameters $\alpha$, $\beta$ in the
quantum Hamiltonian are
shifted by $\pm \frac{1}{8}$.

To conclude, we have two equivalent representations
of the ${\rm P}_{\rm V}$ equation. One is a classical motion in the
time-dependent potential
with Hamiltonian
(\ref{P5aab}). The coordinate of the particle as a function
of time obeys the ${\rm P}_{\rm V}$ equation. Another
representation is a
quantum mechanical particle in the time-dependent
potential of the same form with modified coefficients
described by a non-stationary Schr\"odinger equation.
The latter simultaneously serves as the linear problem
for time evolution associated with the Painlev\'e equation.

\section{Painlev\'e VI}

\subsection{The equation}

The standard form of the ${\rm P}_{\rm VI}$ equation is
\beq\label{P6}
\begin{array}{ll}
\p^2_{T} y= & \displaystyle{
\frac{1}{2}\left (\frac{1}{y}+\frac{1}{y-1}+\frac{1}{y-T}\right )
(\p _{T}y)^2 -\left (\frac{1}{T}+\frac{1}{T-1}+\frac{1}{y-T}\right )
\p_T y}
\\ &\\
& \,\,\, +\,\, \displaystyle{\frac{y(y-1)(y-T)}{T^2 (T-1)^2}
\left ( \alpha +\frac{\beta T}{y^2}+\frac{\gamma (T-1)}{(y-1)^2}
+\frac{\delta T(T-1)}{(y-T)^2}\right ),}
\end{array}
\eeq
where $\alpha$, $\beta , \gamma , \delta$ are arbitrary parameters.
Let us perform the
change of the variables $(y, T) \rightarrow (x,t)$
given by the formulas \cite{Painleve1906,Manin98}
\beq\label{P6a}
y=\frac{\wp (x)-e_1}{e_2 -e_1}\,,
\quad \quad T=\frac{e_3 -e_1}{e_2 -e_1}\,,
\eeq
where $\wp (x)=\wp(x|1, \tau)$ is the Weierstrass
$\wp$-function with periods
$1$, $\tau =2\pi i t$, and $e_k=\wp (\omega _k)$, $k=1,2,3$, are the
values of $\wp (x)$ at the half-periods $\omega _1 =\frac{1}{2}$,
$\omega _2 = \frac{1}{2}(1+\tau )$, $\omega _3 =\frac{1}{2}\tau$.
It is convenient to set also $\omega _0 = 0$.
This change of variables brings the ${\rm P}_{\rm VI}$ equation to the
Newton form
\beq\label{P6b}
\ddot x = \sum_{k=0}^{3}\nu_k \wp ' (x+\omega _k),
\eeq
where
$\wp ' (x)\equiv \p_x \wp (x)$,  and
$\nu_0 =\alpha$, $\nu_1 =-\beta$, $\nu_2 =\gamma$,
$\nu_3 =-\delta +\frac{1}{2}$.
This equation admits the Hamiltonian structure
$$
\dot x =\frac{\p H_{\rm VI}}{\p p}\,, \quad
\quad \dot p =-\frac{\p H_{\rm VI}}{\p x}
$$
with the classical time-dependent Hamiltonian
\beq\label{P6aab}
H_{\rm VI}(p, x)=\frac{p^2}{2}+V_{\rm VI}(x)\,,
\quad \quad
V_{\rm VI}(x)=-\sum_{k=0}^{3}\nu_k \wp (x+\omega _k).
\eeq
It describes classical motion of a point-like particle
in the periodic time-dependent potential.
The time dependence is hidden in the second period
of the $\wp$-function:
\beq\label{ttau}
\wp (x)=\wp(x|\, 1, \tau),\quad \quad \tau = 2\pi i t.
\eeq
To indicate the dependence on the parameters, we will
write $H_{\rm VI}(p, x)=
H^{(\alpha , \beta , \gamma , \delta )}_{\rm VI}(p, x)$
and $V_{\rm VI}(x,t)=
V^{(\alpha , \beta , \gamma , \delta )}_{\rm VI}(x,t)$.
The elliptic form of the ${\rm P}_{\rm VI}$
equation was discussed also in \cite{Babich,Guzzetti}.

\subsection{The zero curvature representation of
the ${\rm P}_{\rm VI}$ equation}

Different versions of the
$U$--$V$ pairs for the ${\rm P}_{\rm VI}$ equation
with spectral parameter on an elliptic curve were
found in \cite{Krichever02} for the special case of
equal constants $\nu_k =\nu$ and in
\cite{Zotov04} for the general case.  However, they appear to be
unsuitable for our purpose. Like in the case of the
${\rm P}_{\rm V}$ equation,
we start from a modified version of the
$U$--$V$ pair with rational dependence on the spectral
parameter suggested in \cite{JM81-2}, then
pass to an elliptic parametrization and
transform it to the desired form by a gauge transformation.

\subsubsection{The modified Jimbo-Miwa $U$--$V$
pair for ${\rm P}_{\rm VI}$}

Let us consider the system of linear problems
\beq\label{P6JM1}
\left \{ \begin{array}{l}
\p_{X}\mathsf{\Psi} ={\sf U}(X,T)\mathsf{\Psi}
\\ \\
\p_T \mathsf{\Psi} ={\sf V}(X,T)\mathsf{\Psi}
\end{array}
\right.
\eeq
with the matrices \cite{JM81-2}
\beq\label{P6JM2}
{\sf U}=\left (
\begin{array}{cc}\displaystyle{\frac{g_0 +\xi_0}{X}
+\frac{g_1 +\xi_1}{X-1}+\frac{g_2 +\xi_2}{X-T}} & \,\,\,\,
\displaystyle{-\! \left (\frac{u_0 g_0}{X}+\frac{u_1 g_1}{X-1}
+\frac{u_2 g_2}{X-T}\right )}
\\ &\\
\displaystyle{\frac{g_0 +2\xi_0}{u_0 X}
+\frac{g_1 +2\xi_1}{u_1(X\! -\! 1)}+
\frac{g_2 +2\xi_2}{u_2(X\! -\! T)}}&\,\,\,\,
\displaystyle{-\! \left (\frac{g_0 +\xi_0}{X}
+\frac{g_1 +\xi_1}{X-1}+\frac{g_2 +\xi_2}{X-T}\right )}
\end{array}\right )
\eeq

\beq\label{P6JM3}
\hspace{-6cm}{\sf V}=\left (
\begin{array}{cc}
\displaystyle{-\, \frac{g_2 +\xi_2}{X-T}} & \,\,\,\,\,\,
\displaystyle{\frac{u_2 g_2}{X-T}}
\\ &\\
\displaystyle{-\, \frac{g_2 +2\xi_2}{u_2 (X-T)}}&\,\,\,\,\,\,
\displaystyle{\frac{g_2 +\xi_2}{X-T}}
\end{array}\right )
\eeq
and two-component vector $\mathsf{\Psi}$.
Here $X$ is the spectral parameter living on the Riemann sphere,
$g_i$, $u_i$ are
some functions of $T$ to be determined
from the zero curvature condition
$\p_X {\sf V}-\p_T {\sf U}+[{\sf V}, \,
{\sf U}]=0$ and $\xi _i$ are arbitrary constants.
Below in this section we denote the entries
of the matrices ${\sf U}$, ${\sf V}$ as
${\sf U}={\sf U}(X)=\left (\begin{array}{ll}{\sf a}(X)&{\sf b}(X)\\
{\sf c}(X)&{\sf d}(X)\end{array}\right )$,
${\sf V}={\sf V}(X)=\left (\begin{array}{ll}{\sf A}(X)&{\sf B}(X)\\
{\sf C}(X)&{\sf D}(X)\end{array}\right )$
(for brevity, the $T$-dependence is not indicated
explicitly).

The following integrals of motion are immediate consequences
of the zero curvature condition:
\beq\label{P6JM4}
\begin{array}{l}
g_0 +g_1 +g_2 =\xi_3
\\ \\
u_0 g_0 + u_1 g_1 + u_2 g_2 =0
\\ \\
\displaystyle{\frac{g_0 +2\xi_0}{u_0}+
\frac{g_1 +2\xi_1}{u_1}+\frac{g_2 +2\xi_2}{u_2}=0.}
\end{array}
\eeq
Here $\xi_3$ is an arbitrary constant, the values of the other two
integrals are set equal to zero following \cite{JM81-2}.
The full system of ordinary differential equations
for the functions $g_i$, $u_i$ which follows from the
zero curvature condition is explicitly given in Appendix A.
Next, let us introduce a function $y$ by representing
the right upper entry of the matrix ${\sf U}$ in the form
\beq\label{P6JM5}
{\sf b}(X)=\frac{K(X-y)}{X(X-1)(X-T)}\,,
\eeq
where
\beq\label{P6JM5b}
K=Tu_0 g_0 + (T-1)u_1 g_1\,,
\quad \quad
y=\frac{Tu_0 g_0}{K}\,.
\eeq
Note that in terms of $K, y$ we have:
\beq\label{P6JM5a}
u_0 g_0 =\frac{Ky}{T}\,, \quad u_1 g_1 =-\, \frac{K(y-1)}{T-1}\,,
\quad u_2 g_2 = \frac{K(y-T)}{T(T-1)}\,.
\eeq
One can see that the zero curvature condition implies
the ${\rm P}_{\rm VI}$ equation (\ref{P6})
for the function $y$ with
\beq\label{P6JM6}
\alpha =2\Bigl (\xi +\frac{1}{2}\Bigr )^2 \,,
\quad
\beta = -2\xi_0^2\,,
\quad
\gamma = 2\xi_1^2 \,,
\quad
\delta =\frac{1}{2}-2\xi_2^2\,,
\eeq
where
\beq\label{P6JM7}
\xi =\xi_0+\xi_1 +\xi_2 +\xi_3\,.
\eeq
Some details of the proof are presented in Appendix A.

\subsubsection{Elliptic parametrization}

The crucial ingredient of the construction is a parametrization
of the modified Jimbo-Miwa $U$--$V$ pair (\ref{P6JM2}), (\ref{P6JM3})
in terms of elliptic functions. This
pa\-ra\-met\-ri\-za\-tion
corresponds to the elliptic substitution (\ref{P6a}) for the dependent
and independent variables leading to the Calogero form
of the ${\rm P}_{\rm VI}$ equation.

We use the general relations
for a change of variables
from $X,T$ to $x,t$ of the form $X=X(x,t)$, $T=T(t)$
given in section 2.4. According
to these relations, the $U$--$V$ pair in the variables $x,t$ is
\beq\label{6p31}
\begin{array}{lll}
{\bf \tilde U} (x,t)& =&\displaystyle{\frac{\p X}{\p x} \,
{\sf U}(X(x,t), T(t))}
\\ &&\\
{\bf \tilde V}(x,t) &=& \displaystyle{\frac{\p T}{\p t} \,
{\sf V}(X(x,t), T(t))
+\frac{\p X}{\p t}\, {\sf U}(X(x,t), T(t)),}
\end{array}
\eeq
where the entries of the matrices ${\sf U}$, ${\sf V}$
in the right hand side should be expressed through the
new variables $x,t$ (see (\ref{6p31})).

We need some formulas which would allow us to make this
transformation explicit. It is natural to expect that the
change of the time variable is the same as for the ${\rm P}_{\rm VI}$
equation itself (see the second formula in (\ref{P6a})).
It turns out that the change of the spectral parameter
is also given by the same elliptic function as the one used for
the dependent variable in (\ref{P6a}):
\beq\label{6p1}
X(x,t)=\frac{\wp (x)-e_1}{e_2 -e_1}\,,
\quad \quad T(t)=\frac{e_3 -e_1}{e_2 -e_1}=\left (
\frac{\vartheta _3(0|\tau )}{\vartheta _0 (0|\tau )}\right )^4.
\eeq
Here
$\wp (x)=\wp (x |1, \tau )$ and $e_j =\wp (\omega_j |1, \tau )$
depend on the new time variable
$t=\frac{\tau}{2\pi i}$ through the second period of the
$\wp$-function. In the last formula, we give the parametrization of
$T$ in terms of Jacobi's theta-functions
$\vartheta _a (x|\tau )$ (see Appendix B).
The arguments leading to equations (\ref{6p1}) and the
derivation are given in Appendix C. Let us also note that
the elliptic substitution
for the spectral parameter of the form (\ref{6p1}) was first
suggested in \cite{LZ06}, where the relation between rational and
elliptic forms of the linear problems for the ${\rm P}_{\rm VI}$
equation was described in terms of modification of the corresponding
vector bundles.

Similar to the previously considered cases,
we deliberately denote the
elliptic spectral parameter by the same
letter $x$ as in the equation of the classical motion (\ref{P6b})
to stress the fact that it is this variable ($x$-coordinate of
a particle on the line) which is going to be
``quantized'' in the ``quantum''
version of the Painlev\'e-Calogero correspondence in the
sense that the momentum
$p=\dot x$ is going to be replaced by the operator $\p_x$.

For practical calculations we need some more formulas.
First of all, we have
\beq\label{6p4}
X=\frac{\wp (x)-e_1}{e_2 -e_1}\,, \quad \quad
X\! - \! 1 =\frac{\wp (x)-e_2}{e_2 -e_1}\,, \quad \quad
X\! - \! T =\frac{\wp (x)-e_3}{e_2 -e_1}\,,
\eeq
so that the identities
\beq\label{6p6}
\left (\frac{\p X}{\p x}\right )^2 =
4(e_2 -e_1)\, X(X-1)(X-T)
\eeq
\beq\label{6p5}
\frac{\p ^2 X}{\p x^2} =
2(e_2 -e_1)\, X(X-1)(X-T)\left ( \frac{1}{X}+
\frac{1}{X-1}+\frac{1}{X-T}\right )
\eeq
hold true. (The first one is the differential
equation for the $\wp$-function (\ref{Bp9}), the second one
is a result of its further differentiating with respect to $x$.)
Next we need the following relations:
\beq\label{6p7}
\begin{array}{rll}
\displaystyle{\frac{(e_2 -e_1)T}{X}}&=&
\wp (x+\omega _1) -e_1
\\ &&\\
\displaystyle{-\, \frac{(e_2 -e_1)(T-1)}{X-1}}&=&
\wp (x+\omega _2) -e_2
\\ &&\\
\displaystyle{\frac{(e_2 -e_1)T(T-1)}{X-T}}&=&
\wp (x+\omega _3) -e_3\,.
\end{array}
\eeq
At last, let us present formulas for derivatives
of the elliptic functions with respect to
$\displaystyle{t=\frac{\tau}{2\pi i}}$.
All of them follow from the ``heat equation''
obeyed by any Jacobi's theta-function
$\vartheta _a (x)=\vartheta _a (x|\tau )$,
$a=0, \ldots , 3$:
\beq\label{heat}
2 \p_t \vartheta _a (x)=\p_x^2 \vartheta _a(x)\,, \quad
t=\frac{\tau}{2\pi i}\,.
\eeq
In particular, we need the following two derivatives:
\beq\label{6p9}
\!\!\! \frac{\p X}{\p t}=\frac{\p X}{\p x}\, \,
\frac{\vartheta _0'(x)}{\vartheta _0(x)}
\eeq
\beq\label{6p8}
\frac{\p T}{\p t}=2(e_2 -e_1) T(T-1).
\eeq
The derivation is given in Appendix B.
The formula for
$\p X /\p t$ first appeared in Takasaki's paper \cite{Takasaki01}.
Note that differentiating a double-periodic function of $x$
with respect to one of the periods, as in (\ref{6p9}),
we obtain a function which is {\it not} an elliptic function of $x$.
The second formula is a direct corollary of the definition
and (\ref{Bp9a}).
(In fact, since $T=X (\omega _3)$, the second formula
follows from the first one).

\subsection{Classical-quantum correspondence for ${\rm P}_{\rm VI}$}

Consider the ${\rm P}_{\rm VI}$ equation in the
Calogero-like form (\ref{P6b}) for a variable $u$:
\beq\label{P6ba}
\ddot u = \sum_{k=0}^{3}\nu_k \wp ' (u +\omega _k).
\eeq
Recall that the variables $u , t$ are connected with the original
variables $y, T$ in (\ref{P6}) by the formulas (\ref{6p1}):
\beq\label{6p10}
y =\frac{\wp (u )-e_1}{e_2 -e_1}\,,
\quad \quad T=\frac{e_3 -e_1}{e_2 -e_1}
\eeq
and
\beq\label{6p11}
\nu _0 = \alpha = 2\Bigl ( \xi +\frac{1}{2}\Bigr )^2,
\quad
\nu _1 =-\beta = 2\xi_0^2, \quad
\nu_2 = \gamma =2\xi_1^2, \quad
\nu_3 = \frac{1}{2}-\delta = 2\xi_2^2\,.
\eeq
This equation is equivalent to the zero curvature
condition for the matrices ${\bf \tilde U}(x,t)$, ${\bf \tilde V}(x,t)$
given by (\ref{6p31}) with the elliptic parametrization (\ref{6p1}).

The next step is a special diagonal
gauge transformation
$\{{\bf \tilde U}, {\bf \tilde V}\}\longrightarrow
\{{\bf U}, {\bf V}\}$ of the form
(\ref{tilde1})
that recasts the matrices in the form
such that the condition $b_x=2B$ (\ref{Bb}) is satisfied.
As is shown in Appendix C, the condition that the dependence
on $x$ and $u$ in the gauge function $\omega$ factorizes
is strong enough
to fix simultaneously the elliptic substitution for the spectral parameter
and the $x$-dependent part of $\omega$.
The latter is found in the form
\beq\label{6cq1}
\omega^2 = \frac{\wp '(x)\, \vartheta _0^2 (x)}{2(\wp (x)\! -\! e_3)}
\, \rho ^2 (t)\,,
\eeq
where $\rho (t)$ is some (yet unknown) function of $t$ only
(see (\ref{s17})).
The function $\rho$ is to be determined at the very end
from the condition that the
$x$-independent part of the potential in the non-stationary
Schr\"odinger equation be equal to the classical
Hamiltonian $H_{\rm VI}(\dot u, u)$.

The detailed derivation of (\ref{6cq1}) is given in Appendix C.
Here we can only say that if this expression is known, then
it is an easy exercise to check that the
$x$-derivative of the right upper element $b$
of the matrix ${\bf U}$,
\beq\label{6cq2}
b = \omega^2 \, \tilde b =
2K (e_2-e_1)\rho^2(t) \vartheta_0^2 (x)\, \frac{\wp (x)-\wp (u )}{\wp (x)-e_3}\,,
\eeq
appears to be
equal to $2B = 2 \omega ^2 \tilde B$.
Therefore, in this gauge the non-stationary
Schr\"o\-din\-ger
equation of the form (\ref{P1eea}) does hold. It remains to
find the potential $U(x,t)= \frac{1}{2}(ad -bc -a_x +2A)$.
Taking into account that $a=\tilde a +\p_x \log \omega$,
$A =\tilde A +\p_t \log \omega$ and
$\displaystyle{\tilde a =  \frac{\p X}{\p x}\, {\sf a}}$,
$\displaystyle{\tilde A =  \frac{\p T}{\p t}\, {\sf A}
+\frac{\p X}{\p x}\, \frac{\vartheta _0'(x)}{\vartheta _0(x)}\,
{\sf a}}$,
one can represent it as a sum of three terms:
$U=U_1 + U_2 +U_3$, where
$$
\begin{array}{lll}
U_1 &=&
\displaystyle{\frac{1}{2} \left (\frac{\p X}{\p x}\right )^2
\det {\sf U}(X) \, =\,
\frac{1}{2} \left (\frac{\p X}{\p x}\right )^2
({\sf a}{\sf d}-{\sf b}{\sf c})}
\\ && \\
U_2& =& \displaystyle{-\frac{1}{2}\left [\frac{\p^2 X}{\p x^2}\,
{\sf a} + \left (\frac{\p X}{\p x}\right )^2 {\sf a}_{X}\right ]
+\frac{\p T}{\p t}\, {\sf A}
+\frac{\p X}{\p x}\, \frac{\vartheta _0'(x)}{\vartheta _0(x)}\,
{\sf a} -\frac{\p X}{\p x}\, {\sf a} \, \p_x \log \omega}
\\ && \\
U_3& =& \displaystyle{-\frac{1}{2}(\p_x \log \omega )^2 -
\frac{1}{2}\, \p_x^2 \log \omega +\p_t \log \omega}.
\end{array}
$$

For the purpose of this paper we do not need the
explicit form of the matrices ${\bf U}(x,t)$, ${\bf V}(x,t)$
in the elliptic parametrization (this will be presented elsewhere).
Technically, it is convenient to make the calculations using the
original variables $X,T$ where possible and pass
to the elliptic parametrization at the very end. That is why
we have expressed the right hand sides in terms of the
matrices ${\sf U}$, ${\sf V}$ with rational dependence on the
spectral parameter $X$.

The calculation of the $X$-dependent part of
$U_1$ is relatively easy.
The result is:
\beq\label{P6cq1}
\begin{array}{lll}
U_1 & =&\displaystyle{
2(e_2 -e_1)\left [ -\, \xi^2 X -\frac{T\xi_0^2}{X}
+\frac{(T-1)\xi_1^2}{X-1}-\frac{T(T-1)\xi_2^2}{X-T}\right ]
+U_{1,0}}
\\ &&\\
&=&
\displaystyle{-\, 2\xi ^2 (\wp (x)-e_1)-2\xi_0^2 (\wp (x+\omega_1)-e_1)
-2\xi_1^2 (\wp (x+\omega_2)-e_2)}
\\ &&\\
&&\displaystyle{
-\,\, 2\xi_2^2 (\wp (x+\omega_3)-e_3)+
U_{1,0}\,.}
\end{array}
\eeq
The passage to the elliptic functions is done according to formulas
(\ref{6p7}).
The $X$-independent part, $U_{1,0}$, is
\beq\label{P6cq2}
U_{1,0} =
2\xi (e_2 -e_1)\Bigl [ (T+1)(g_0 +\xi _0)
 +(T-1) (g_1 +\xi _1) - (T-1) (g_2 +\xi _2) \Bigr ].
\eeq
Using formulas from Appendix A, we get:
\beq\label{P6cq3}
\begin{array}{lll}
U_{1,0}& =&\displaystyle{
\Bigl (2\xi^2 -\frac{1}{2}\Bigr )(e_2 -e_1)y
\, +\, \frac{2(e_2 -e_1)T}{y}\,
\left ( -\frac{(T-1)^2 y_T^2}{4} +\xi_0^2\right )}
\\ &&\\
&&\hspace{-1cm} +\,\,\,
\displaystyle{\frac{2(e_2 \! -\! e_1)(T\! -\! 1)}{y-1}\,
\left (\frac{T^2y_T^2}{4}  -\xi_1^2\right )
+\frac{2(e_2 \! -\! e_1)T(T\! -\! 1)}{y-T}\,
\left (-\frac{(y_T\! -\! 1)^2}{4}  +\xi_2^2 \right )}
\\ &&\\
&&\hspace{-1cm} -\,\,\,
\displaystyle{
\frac{(e_2\! -\! e_1)(T\! -\! 1)}{2}\,.}
\end{array}
\eeq
The $T$-derivative of $y$ in the elliptic parametrization reads
\beq\label{P6cq4}
y_T = \frac{1}{2T(T\! -\! 1)}\,
\frac{\wp '(u )}{(e_2 \! -\! e_1)^2}
\left (\dot u + \frac{\vartheta_0'(u )}{\vartheta_0(u )}
\right ).
\eeq
Plugging it to the right hand side of (\ref{P6cq3}) and using
formulas (\ref{6p7}) (now with $y, u$ instead of $X, x$),
we obtain:
\beq\label{P6cq5}
\begin{array}{lll}
U_{1,0}& =&\displaystyle{
\Bigl (2\xi^2 -\frac{1}{2}\Bigr )(\wp (u )-e_1)
+2\xi_0^2 (\wp (u +\omega_1) -e_1)}
\\ &&\\
&&\displaystyle{+\,\,
2\xi_1^2 (\wp (u +\omega_2) -e_2)+
\Bigl (2\xi_2^2 -\frac{1}{2}\Bigr )(\wp (u +\omega_3) -e_3)}
\\ &&\\
&&\displaystyle{-\,\, \frac{1}{2}\left (\dot u +
\frac{\vartheta_0'(u )}{\vartheta_0 (u )}\right )^2
+ \frac{\wp '(u )}{2(e_2 \! -\! e_1)(y\! -\! T)}\,
\left (\dot u +
\frac{\vartheta_0'(u )}{\vartheta_0 (u )}\right )
-\frac{e_3-e_2}{2}\,.}
\end{array}
\eeq
The unwanted terms in the last line can be transformed to
logarithmic $t$-derivatives using the formulas
\beq\label{P6cq5a}
\p_t \log \vartheta _0(u )=
\dot u \, \frac{\vartheta _0'(u )}{\vartheta _0(u )}
+\frac{1}{2}
\left (\frac{\vartheta _0'(u )}{\vartheta _0(u )}\right )^2
-\frac{1}{2}\, \wp (u +\omega _3)-\eta
\eeq
\beq\label{P6cq5b}
\frac{1}{2}\,
\p_t \log (y-T)=\frac{\wp '(u )}{2(e_2 \! -\! e_1)(y\! -\! T)}\,
\left (\dot u +
\frac{\vartheta_0'(u )}{\vartheta_0 (u )}\right )
-\wp (u +\omega _3)+e_3
\eeq
(and thus they can be eliminated by a proper choice of
the function $\rho (t)$, see below). Taking this into account,
we obtain $U_{1,0}$ in the form
\beq\label{P6cq5c}
\begin{array}{lll}
U_{1,0}& =&\displaystyle{
\Bigl (2\xi^2 -\frac{1}{2}\Bigr )(\wp (u )-e_1)
+2\xi_0^2 (\wp (u +\omega_1) -e_1)}
\\ &&\\
&&\displaystyle{+\,\,
2\xi_1^2 (\wp (u +\omega_2) -e_2)+
2\xi_2^2 (\wp (u +\omega_3) -e_3)}
\\ &&\\
&&\displaystyle{-\,\, \frac{\dot u^2}{2}+\p_t \log
\frac{(y-T)^{1/2}}{\vartheta_0(u )}-e_3+\frac{e_2}{2}-\eta}.
\end{array}
\eeq

For the calculation of $U_2$ we prepare the formulas
$$
\begin{array}{lll}
\displaystyle{\frac{\p X}{\p x}\, \p_x \log \omega}
&=&\displaystyle{\frac{1}{2}\, \frac{\p X}{\p x}\,
\p_x \log \left ( \frac{\wp '(x)}{e_2 -e_1}\,
\frac{e_2 -e_1}{\wp (x)-e_3}\right ) +
\frac{\p X}{\p x}\, \frac{\vartheta_0'(x)}{\vartheta_0 (x)}}
\\ &&\\
&=&\displaystyle{\frac{1}{2}\left [\frac{\p^2 X}{\p x^2}-
\frac{1}{X-T}\, \Bigl ( \frac{\p X}{\p x} \Bigr )^2 \right ]
+\frac{\p X}{\p x}\, \frac{\vartheta_0'(x)}{\vartheta_0 (x)}}
\end{array}
$$
$$
\frac{\p^2 X}{\p x^2}\, {\sf a}+\frac{1}{2}
\Bigl ( \frac{\p X}{\p x} \Bigr )^2\, {\sf a}_X
=2(e_2-e_1) \Bigl [ 2\xi X -(T+1) (g_0 \! +\! \xi_0)
-T (g_1\! +\! \xi_1) \! -\! (g_2 \! +\! \xi_2)\Bigr ]
$$
and
$$
\frac{1}{2}
\Bigl ( \frac{\p X}{\p x} \Bigr )^2 \frac{{\sf a}}{X-T}
+\frac{\p T}{\p t}\, {\sf A} =
2(e_2-e_1) \Bigl [ \xi X -(g_0 +\xi_0)+(T-1)(g_2 +\xi_2)\Bigr ].
$$
The calculation gives the following simple result:
\beq\label{P6cq6}
U_2=-2\xi (\wp (x)-e_3).
\eeq

At last, let us find $U_3$.
We have:
$$
\begin{array}{lll}
\p_t \log \omega &=&
\displaystyle{\frac{1}{2}\, \p_t \log \left (
\frac{\wp '(x)}{\wp (x)-e_3}\right )+\p_t \log (\rho \vartheta_0 (x))}
\\ &&\\
&=&\displaystyle{\frac{1}{2}\, \p_t \log\frac{\p X}{\p x}
-\frac{1}{2}\,\p_t \log (X-T)+\p_t \log (\rho \vartheta_0 (x))}
\\ &&\\
&=&\displaystyle{\frac{1}{2}\,\left (\frac{\p X}{\p x}\right )^{-1}
\frac{\p^2 X}{\p x^2}
+\frac{1/2}{X-T}\left ( \frac{\p T}{\p t}-\frac{\p X}{\p t}\right )
+\p_t \log \vartheta_0 (x)  +\p_t \log \rho}
\\ &&\\
&=&\displaystyle{\frac{1}{2}\, \p_x \log
\left (\frac{\wp '(x)}{\wp (x)-e_3}\right )
\p_x \log \vartheta_0 (x) +\frac{1}{2}\, (\p_x \log \vartheta_0 (x))^2
+\p_t \log \rho -e_3 -2\eta}\,,
\end{array}
$$
where $\displaystyle{\eta =-\frac{1}{6} \,
\frac{\vartheta_1^{'''}(0)}{\vartheta_1 (0)}}$. When passing to the
last line we have used the heat equation (\ref{heat}) and
the relation $\p_x^2 \log \vartheta_0 (x)=-\wp (x+\omega_3)
-2\eta$. Combining the different contributions to $U_3$
and passing
to the elliptic parametrization, we find:
\beq\label{P6cq7}
U_3 = -\frac{1}{8}\Bigl ( 3\wp (x)-\wp (x+\omega_1)
-\wp (x+\omega_2)-\wp (x+\omega_3)\Bigr ) -\frac{e_3}{2}-\eta
+\p_t \log \rho \,.
\eeq

Using the formulas given above and equation (\ref{Ap8}),
we obtain the potential
in the form
\beq\label{P6cq8}
\begin{array}{lll}
U(x,t)&=&\displaystyle{-\left ( 2\Bigl ( \xi \! +\!
\frac{1}{2}\Bigr )^2 \! -\! \frac{1}{8}\right )
\wp (x)-\left (2\xi_0^2 -\frac{1}{8}\right )\wp (x+\omega_1)}
\\ &&\\
&&\displaystyle{\,\, -\,\,
\left (2\xi_1^2 -\frac{1}{8}\right )\wp (x+\omega_2)
-\left (2\xi_2^2 -\frac{1}{8}\right )\wp (x+\omega_3)\! -\! \tilde H}\,,
\end{array}
\eeq
where
$$
\tilde H =\frac{\dot u ^2}{2}
-\sum_{k=0}^{3}\nu_k \wp (u +\omega_k)
+\frac{1}{2}\, \p_t \log \left (\frac{\vartheta_{0}^{2}(u )
(\vartheta _{1}'(0))^{\frac{8}{3}}}{(y-T)K(T)
\vartheta_{0}^{6}(0)\rho^2(t)}\right )
$$
with the same $\nu_k$ as in (\ref{6p11}).
Using the identities from Appendix B one can express
$y-T$ in terms of the theta-functions:
$$
\frac{1}{y-T}= \frac{e_2-e_1}{\wp (u )-e_3}
= -\, \frac{\pi^2
\vartheta_{0}^{6}(0)
\vartheta_{1}^{2}(u )}{(\vartheta _{1}'(0))^{2}
\vartheta_{0}^{2}(u )}\,.
$$
Therefore, choosing
$$
\rho (t)=
\frac{(\vartheta _{1}'(0))^{\frac{1}{3}}
\, \vartheta_1(u )}{\sqrt{K(T)}}\,,
$$
we see that
$\tilde H=H_{\rm VI}^{(\alpha , \beta ,
\gamma , \delta )}(\dot u , u )$ with the same
parameters as in (\ref{P6JM6}). With this choice of
$\rho$, the gauge function $\omega$ (\ref{6cq1}) acquires the form
\beq\label{P6cq9}
\omega^2 = \frac{(\vartheta _{1}'(0))^{\frac{5}{3}}\,
\vartheta_0 (0)}{
\vartheta_2 (0)\vartheta_3 (0)}\,
\frac{\vartheta_2 (x)\vartheta_3 (x)\vartheta_0 (x)
\vartheta_1^2 (u )}{\vartheta_1 (x)K(T)}\,.
\eeq

Finally, we conclude that the classical-quantum correspondence
does work for the ${\rm P}_{\rm VI}$ equation. The non-stationary
Schr\"odinger equation is
\beq\label{P6cq10}
\p_t \psi =\Bigl [H_{\rm VI}^{(\alpha -\frac{1}{8}\, ,
\beta +\frac{1}{8}\, , \gamma -\frac{1}{8}\, ,
\delta +\frac{1}{8})}(\p_x , x)\, - \,
H_{\rm VI}^{(\alpha , \beta , \gamma ,
\delta )}(\dot u , u )\Bigr ]\psi ,
\eeq
where
$$
H_{\rm VI}^{(\alpha -\frac{1}{8}\, ,
\beta +\frac{1}{8}\, , \gamma -\frac{1}{8}\, ,
\delta +\frac{1}{8})}(\p_x , x)=
\frac{1}{2}\, \p_x^2 -\sum_{k=0}^{3}\Bigl (\nu_k -
\frac{1}{8}\Bigr )
\wp (x+\omega _k)
$$
and the parameters $\nu_k$ are connected with $\alpha , \beta ,
\gamma , \delta$ as in (\ref{6p11}). The function
\beq\label{P6i}
\Psi (x,t)=e^{\int^t H_{\rm VI}^{(\alpha , \, \beta , \, \gamma ,\,
\delta )} (\dot u , u)dt'}\psi (x,t) \eeq
thus obeys the
non-stationary Schr\"odinger equation
\beq\label{P6j}
\p_t \Psi =
H_{\rm VI}^{(\alpha -\frac{1}{8}, \, \beta +\frac{1}{8}, \, \gamma
-\frac{1}{8},\, \delta +\frac{1}{8} )} (\p_x , x)\Psi =\left (
\frac{1}{2}\, \p_x^2 + V_{\rm VI}^{(\alpha -\frac{1}{8}, \, \beta
+\frac{1}{8}, \, \gamma -\frac{1}{8},\, \delta +\frac{1}{8} )}
(x,t)\right )\Psi .
\eeq
Note that in the quantum part all the
parameters undergo shifts by $\pm \frac{1}{8}$. In terms of the
parameters $\nu_k$ (see (\ref{P6b}), (\ref{6p11})) the shifts are
$\nu_k \rightarrow \nu_k -\frac{1}{8}$, $k=0, \ldots , 3$. In
particular, if all $\nu_k$ are equal to each other,
$\nu_k =\nu$, then we obtain the
non-stationary Lam\'e equation
\beq\label{P5l}
\p_t \Psi =\left
( \frac{1}{2}\, \p_x^2 -4\tilde \nu \wp \Bigl (2x\, \Bigl |\,  1,
2\pi i t \right )\Psi\,, \quad \tilde \nu =\nu
-\frac{1}{8}\,.
\eeq
(the identity $\sum\limits_{k=0}^{3}\wp (x+\omega _k)=4\wp (2x)$
has been used).
We remark that the non-stationary Lam\'e equation in connection with
the ${\rm P}_{\rm VI}$ equation (and with the
8-vertex model) was discussed in \cite{BM06}.
Recently, the non-stationary Lam\'e equation has appeared
\cite{FL,MMM} in the context of the AGT conjecture.

To summarize, similar to the other cases,
we have two equivalent representations
of the ${\rm P}_{\rm VI}$ equation. One is a classical motion in the
time-dependent periodic potential
with Hamiltonian
(\ref{P6aab}). The coordinate of the particle as a function
of time obeys the ${\rm P}_{\rm VI}$ equation. Another
representation is a
quantum mechanical particle in the time-dependent
potential of the same form with modified coefficients
described by a non-stationary Schr\"odinger equation.
The latter simultaneously serves as the linear problem
for time evolution associated with the Painlev\'e equation.

\section{Concluding remarks}

We have shown that
for each Painlev\'e equation written in the ``Calogero
form'' $\ddot u =- \p_u V(u,t)$
with a time-dependent potential $V(x,t)$, the associated
linear problems can be represented as
\beq\label{scalareqs2}
\left \{
\begin{array}{l}\displaystyle{
\left (\frac{1}{2}\, \p_{x}^{2} -\frac{1}{2}\,
(\p_x \log b (x,t)) \, \p_x \,  + \tilde W (x,t)\right )\Psi =E\Psi}
\\ \\
\displaystyle{
\p_t \Psi =\left ( \frac{1}{2}\, \p_x^2 +\tilde V(x,t)\right )\Psi},
\end{array}
\right.
\eeq
The second equation is the
{\it non-stationary Schr\"odinger equation in imaginary time
with the potential $\tilde V(x,t)$ that
has the same form as
the classical potential for the Painlev\'e equation}
(with possibly modified parameters).
The potential in the first equation is
$$
\tilde W (x,t)=\tilde V(x,t)-\frac{\p_t b(x,t)}{2b(x,t)}
+\frac{\p_x^2 b(x,t)}{4b(x,t)}
$$
and the eigenvalue $E$ is the value
of the classical Hamiltonian $H(\dot u, u)$ for the
Painlev\'e equation in the Calogero form (with the opposite sign):
$$
E= - H(\dot u, u)=-\frac{\dot u^2}{2}-V(u,t).
$$
These equations has been derived from the $2\times2$ matrix
linear problems (\ref{d001}) with the matrices
${\bf U}(x,t)$, ${\bf V}(x,t)$ of the special form,
with the function
$b(x,t)$ being the right upper entry of the matrix
${\bf U}(x,t)$.
The second equation of the system
(\ref{scalareqs2}) describes isomonodromic deformations
of the first one and
their compatibility implies the Painlev\'e equation (in the Calogero
form) for the function
$u=u(t)$ defined implicitly as  zero of the function $b(x,t)$:
$b(u(t), t)=0$.

In short, the conclusion is that linearization
of the Painlev\'e equation,
i.e., passing to the linear problem,
is equivalent to its quantization.
The imaginary
time suggests an interpretation in terms of the Fokker-Planck
equation for a stochastic process.

Here a remark is in order. On the one hand, the Painlev\'e equation
is obtained as a compatibility condition for the pair of equations
(\ref{scalareqs2}). However, on the other hand,
the second equation alone is already enough
to encode the full information about the Painlev\'e equation. Indeed,
it describes a quantum mechanical particle on the line
in the time-dependent potential corresponding to the Painlev\'e
equation. Therefore, the Painlev\'e equation itself should emerge
in the classical limit.

In the papers \cite{Reshet92,Harnad94}
the Knizhnik-Zamolodchikov system of
equations was treated as a natural
quantization of isomonodromic deformations.
It would be very interesting to understand our results
in these terms.

At last, we would like to point out that another sort of
classical-quantum correspondence for Painlev\'e equations
was established in the work \cite{Slavyanov}. Namely, it was shown
that each equation from the Painlev\'e list
could be regarded as a ``classical analog''
of a linear ordinary differential equation
of the Heun class in the sense that
the second-order differential operator $\hat L (\p_x , x)$
involved in the latter, after a
properly taken classical limit, coincides with
the polynomial classical Hamiltonian for the Painlev\'e
equation. (In other words, the Euler-Lagrange equation
corresponding to the symbol $L(p,q)$
of the linear differential operator $\hat L$ is just the
Painlev\'e equation.) Similarly to our approach,
in this classical/quantum mechanical
interpretation, the time variable $T$ has the meaning of the
deformation parameter. However, the
important difference is that \cite{Slavyanov}
deals with {\it stationary}
Schr\"odinger-like equation with coefficients depending on $T$.
It seems to us that
the construction elaborated in the present paper
and a similar construction suggested previously in \cite{Suleimanov94}
are more appropriate
because the Painlev\'e equations are essentially non-autonomous
systems and it is really natural to associate
{\it non-stationary} Schr\"odinger equations with them.

\section*{Acknowledgments}
The authors are grateful to I.Krichever, S.Ob\-le\-zin
and V.Poberezhniy for discussions. They also thank V.Po\-be\-rezh\-niy
and B.Su\-lei\-ma\-nov for bringing the papers
\cite{Slavyanov} and \cite{Suleimanov94,Suleimanov08,DNovikov09}
to their attention.
The work of both authors was partially supported by Russian
Federal Nuclear Energy Agency under contract H.4e.45.90.11.1059.
The work of
A.Zabrodin was supported in part by RFBR grant 11-02-01220, by joint
RFBR grants 09-01-93106-CNRS, 10-02-92109-JSPS
and by the Federal Agency for Science and Innovations
of Russian Federation under contract 14.740.11.0081.
The work of
A.Zotov was supported in part by grants RFBR-09-02-00393,
RFBR-09-01-92437-KEa, RFBR-09-01-93106-CNRS, Russian President fund
MK-1646.2011.1 and by the Federal Agency for Science and Innovations
of Russian Federation under contract 14.740.11.0347.

\section*{Appendix A}
\addcontentsline{toc}{section}{Appendix A}
\def\theequation{A\arabic{equation}}
\setcounter{equation}{0}

In this appendix we present some details of the derivation
of the ${\rm P}_{\rm VI}$ equation from the zero curvature condition
for matrices (\ref{P6JM2}), (\ref{P6JM3}). We use the notation introduced
in the main text.

First of all, let us write down the differential equations
for the functions $g_i$, $u_i$ that follow from the zero
curvature condition. The full system of equations reads
\beq\label{Ap0}
\begin{array}{rll}
T\p_T (u_0 g_0) & =& 2 u_0 g_0 (g_0 +g_2 +\xi _0 +\xi _2)
+2 u_1 g_1 (g_0 + \xi _0)
\\ && \\
(T-1)T\p_T (u_1 g_1) & =& 2 u_1 g_1 (g_1 +g_2 +\xi _1 +\xi _2)
+2 u_0 g_0 (g_1 + \xi _1)
\\ && \\
T\p_T g_0 & = & \displaystyle{
\frac{u_0}{u_2}\, g_0 (g_2 +2\xi _2) -
\frac{u_2}{u_0}\, g_2 (g_0 +2\xi _0)}
\\ && \\
(T-1)\p_T g_1 & = & \displaystyle{
\frac{u_1}{u_2}\, g_1 (g_2 +2\xi _2) -
\frac{u_2}{u_1}\, g_2 (g_1 +2\xi _1)}
\\ && \\
\displaystyle{T \p_T \Bigl (
\frac{g_0 +2 \xi _0)}{u_0}\Bigr )} & = &
\displaystyle{ \frac{2}{u_0}(g_0 +2\xi _0)
(g_2 + \xi _2)-\frac{2}{u_2}(g_2 +2\xi _2)
(g_0 + \xi _0)}
\\ &&\\
\displaystyle{(T-1)\p_T \Bigl (
\frac{g_1 +2 \xi _1)}{u_1}\Bigr )} & = &
\displaystyle{ \frac{2}{u_1}(g_1 +2\xi _1)
(g_2 + \xi _2)-\frac{2}{u_2}(g_2 +2\xi _2)
(g_1 + \xi _1)}
\end{array}
\eeq
along with the integrated relations (\ref{P6JM4}).
However, a direct derivation of
the  ${\rm P}_{\rm VI}$ equation
from this system is not the easiest way. Below
we give a short-cut which closely follows the derivation
outlined in \cite{JM81-2}.

Along with the function $y$ defined by (\ref{P6JM5})
let us also introduce the function
\beq\label{Ap1}
z={\sf a}(y)=\frac{g_0 +\xi_0}{y}+\frac{g_1 +\xi_1}{y-1}+
\frac{g_2 +\xi_2}{y-T}\,.
\eeq
Then, from the fact that the total $T$-derivative of
${\sf b}(y)$ is zero, we write, using the zero curvature
equations in the form (\ref{zc1}):
$0=d{\sf b}(y)/dT = {\sf b}_X (y) y_T +{\sf b}_T(y)=
{\sf b}_T (y) y_T + {\sf B}_X(y)-2z {\sf B}(y)=0$,
where $y_T \equiv dy/dT$.
Expressing ${\sf b}_X (y)$, etc in terms of
the functions $K$ and $y$ (see (\ref{P6JM5b})),
we obtain:
\beq\label{Ap2}
y_T = \frac{y(y-1)(y-T)}{T(T-1)}\, \Bigl (
2z +\frac{1}{y-T}\Bigr ).
\eeq

Combining the integrals of motion (\ref{P6JM4}) with the
definition of $z$, and using formulas (\ref{P6JM5a}),
we can write the system of equations
\beq\label{Ap3}
\left \{\begin{array}{l}
g_0 +g_1 +g_2 =\xi_3
\\ \\
\displaystyle{\frac{g_0 +\xi_0}{y}+\frac{g_1 +\xi_1}{y-1}+
\frac{g_2 +\xi_2}{y-T} =z}
\\ \\
\displaystyle{\frac{Tg_0(g_0 +2\xi_0)}{y}-
\frac{(T-1)g_1 (g_1 +2\xi_1 )}{y-1}+
\frac{T(T-1)g_2 (g_2 +2\xi_2 )}{y-T}=0}
\end{array}\right.
\eeq
for the three functions $g_i$ which can be solved as

\beq\label{Ap40}
\begin{array}{ll}
g_0 =&\displaystyle{-\, \frac{y}{2\xi T}\left [
y(y-1)(y-T)\tilde z^2 -2\Bigl ( \xi_3 (y-1)(y-T)
-\xi_1 (y-T) -\xi_2 T(y-1)\Bigr )\tilde z\right.}
\\ & \\
&\left. \phantom{int}\,\,\, + \, \xi _3
\Bigl ( \xi_3 (y-1)-(2\xi_2 +\xi_3)T-2\xi_1\Bigr )\right ]
\end{array}
\eeq

\beq\label{Ap41}
\begin{array}{ll}
\hspace{-0.3cm}g_1 =&\displaystyle{\frac{y-1}{2\xi (T-1)}\left [
y(y-1)(y-T)\tilde z^2 -2\Bigl ( \xi_3 y(y-T)
+\xi_0 (y-T) -\xi_2 (T-1)y\Bigr )\tilde z\right.}
\\ & \\
&\left. \phantom{int}\,\,\, + \, \xi_3
\Bigl ( \xi_3 (y-1)-(2\xi_2 +\xi_3)(T-1)+2\xi_0 +\xi_3\Bigr )\right ]
\end{array}
\eeq

\beq\label{Ap42}
\begin{array}{ll}
g_2 =&\displaystyle{-\, \frac{y-T}{2\xi T(T-1)}\left [
y(y-1)(y-T)\tilde z^2 -2\Bigl ( \xi_3 y(y-1)
+\xi_0 T(y-1) +\xi_1 (T-1)y\Bigr )\tilde z\right.}
\\ & \\
&\left. \phantom{int}\,\,\, + \, \xi_3
\Bigl ( \xi_3 (y-1)+(2\xi_0 +\xi_3)T+2\xi_1 (T-1)\Bigr )\right ],
\end{array}
\eeq
where
\beq\label{Ap5}
\tilde z = z-\frac{\xi_0}{y}-\frac{\xi_1}{y-1}-\frac{\xi_2}{y-T}
\,\, =\,\, \frac{g_0}{y}+\frac{g_1}{y-1}+\frac{g_2}{y-T}
\eeq
and $\xi = \xi_0+\xi_1 +\xi_2 +\xi_3 $. In order to find a
more explicit representation, we notice that the functions
$g_0 /y$, $g_1/(y-1)$ and $g_2/(y-T)$ are rational functions of
the variable $y$ with first order poles at $0$, $1$, $T$ and $\infty$.
Calculating the residues, one can write them in the explicit form:
\beq\label{Ap60}
\begin{array}{ll}
\displaystyle{2\xi \, \frac{g_0}{y}}
=&\displaystyle{-\, \frac{(\xi +\frac{1}{2})^2}{T}\, y -
\frac{G_0}{T} -\frac{1}{y}\left [ \frac{(T-1)^2}{4}y_T^2 -
\xi (T-1) y_T +\xi_0 (2\xi -\xi_0)\right ]}
\\ & \\
&\displaystyle{ \,\,\, + \,\frac{T-1}{T(y-1)}
\left [ \frac{T^2}{4}y_T^2 -\xi_1^2 \right ]
\, -\, \frac{T-1}{y-T}\left [ \frac{1}{4}(y_T -1)^2 -\xi_2^2
\right ]}
\end{array}
\eeq

\beq\label{Ap61}
\begin{array}{ll}
\displaystyle{2\xi \, \frac{g_1}{y-1}}
=&\displaystyle{ \frac{(\xi +\frac{1}{2})^2}{T-1}\, y +
\frac{G_1}{T-1} +\frac{T}{(T\! -\! 1)y}
\left [ \frac{(T-1)^2}{4}y_T^2 -
\xi_0^2\right ]}
\\ & \\
&\displaystyle{ \!\!\!\! + \, \frac{1}{y-1}
\left [-\, \frac{T^2}{4}y_T^2
-\xi T y_T +\xi_1 (\xi_1 \! -\! 2\xi )\right ]+
\frac{T}{y\! -\! T}
\left [ \frac{1}{4}(y_T  -\! 1)^2 -\xi_2^2
\right ]}
\end{array}
\eeq

\beq\label{Ap62}
\begin{array}{ll}
\displaystyle{2\xi \, \frac{g_2}{y-T}}
=&\displaystyle{-\,\, \frac{(\xi +\frac{1}{2})^2}{T(T\! -\! 1)}\, y
-\frac{G_2}{T(T\! -\! 1)}-\frac{1}{(T\! -\! 1)y}
\left [ \frac{(T-1)^2}{4}y_T^2 -\xi_0^2 \right ]}
\\ & \\
&\!\!\!\! \displaystyle{  + \, \frac{1}{T(y-1)}\left
[\frac{T^2}{4}y_T^2 \! -\! \xi_1^2 \right ]\, -\, \frac{1}{y-T}\left
[ \frac{1}{4}(y_T -1)^2 -\xi  (y_T-1) +\xi_2 (2\xi \! -\! \xi_2)
\right ]}
\end{array}
\eeq
and
\beq\label{Ap7}
\begin{array}{ll}
G_0 =& \displaystyle{\frac{T\! -\! 1}{4}-\xi (\xi T +\xi +1)}
\\ &\\
G_1 =& \displaystyle{\frac{T\! -\! 1}{4}-\xi ^2 (T-1)}
\\ &\\
G_2 =& \displaystyle{\frac{T\! -\! 1}{4}+\xi (\xi +1) (T-1).}
\end{array}
\eeq

The next step is to express the $T$-derivative of the
function $z$ in terms of the functions $g_i$ and $y$.
For that purpose, we write $ z_T = {\sf a}_{X}(y) y_T
+{\sf a}_{T}(y)$ and use the zero curvature equation
${\sf a}_{T}(X)-{\sf A}_{X}(X)+{\sf b}(X){\sf C}(X)
-{\sf c}(X){\sf B}(X)=0$
to obtain
$$
\begin{array}{lll}
z_T &=& {\sf a}_{X}(y) y_T + {\sf A}_{X}(y)
+{\sf c}(y){\sf B}(y)
\\ &&\\
&=&\,\,\, - \displaystyle{
\left (\frac{g_0 +\xi_0 }{y^2} +\frac{g_1 +\xi_1 }{(y-1)^2}
+\frac{g_2 +\xi_2 }{(y-T)^2}\right ) y_T \, +\,
\frac{g_2 +\xi_2) }{(y-T)^2}}
\\ &&\\
&& \,\,\, +\, \displaystyle{\frac{1}{T(T-1)} \left ( \frac{Tg_0(g_0
+2\xi_0 )}{y^2} - \frac{(T\! -\! 1)g_1 (g_1 +2\xi_1) }{(y-1)^2}
+\frac{T(T\! -\! 1)g_2 (g_2 +2\xi_2) }{(y-T)^2}\right ).}
\end{array}
$$
It remains to express $z_T$ in terms of $y$, $y_T$,
$y_{TT}$ with the help of (\ref{Ap2}) and to plug the explicit form
of the functions $g_i$ given by equations (\ref{Ap60})--(\ref{Ap62}).
After a long calculation, one obtains the ${\rm P}_{\rm VI}$
equation (\ref{P6}) with the parameters (\ref{P6JM6}).

In establishing the classical-quantum correspondence
we also need the $T$-derivative of the function $K(T)$.
A straightforward calculation, which uses formulas
(\ref{P6JM5a}) and the first two equations of the system
(\ref{Ap0}), yields:
\beq\label{Ap8}
\p_T \log K = -(2\xi +1) \frac{y-T}{T(T\! -\! 1)}\,.
\eeq

\section*{Appendix B}
\addcontentsline{toc}{section}{Appendix B}
\def\theequation{B\arabic{equation}}
\setcounter{equation}{0}

\subsection*{Theta-functions, Weierstrass $\wp$-function
and other useful functions}

\paragraph{Theta-functions.}
The Jacobi's theta-functions $\vartheta_a (z)=
\vartheta_a (z|\tau )$, $a=0,1,2,3$, are defined by the formulas
\beq\label{Bp1}
\begin{array}{l}
\vartheta _1(z)=-\displaystyle{\sum _{k\in \z}}
\exp \left (
\pi i \tau (k+\frac{1}{2})^2 +2\pi i
(z+\frac{1}{2})(k+\frac{1}{2})\right ),
\\ \\
\vartheta _2(z)=\displaystyle{\sum _{k\in \z}}
\exp \left (
\pi i \tau (k+\frac{1}{2})^2 +2\pi i
z(k+\frac{1}{2})\right ),
\\ \\
\vartheta _3(z)=\displaystyle{\sum _{k\in \z}}
\exp \left (
\pi i \tau k^2 +2\pi i
z k \right ),
\\ \\
\vartheta _0(z)=\displaystyle{\sum _{k\in \z}}
\exp \left (
\pi i \tau k^2 +2\pi i
(z+\frac{1}{2})k\right ),
\end{array}
\eeq
where $\tau$ is a complex parameter
(the modular parameter) such that ${\rm Im}\, \tau >0$.
The function
$\vartheta _1(z)$ is odd, the other three functions are even.
The infinite product representation for the $\vartheta_1(z)$
reads:
\beq
\label{infprod}
\vartheta_1(z)=i\,\mbox{exp}\, \Bigl (
\frac{i\pi \tau}{4}-i\pi z\Bigr )
\prod_{k=1}^{\infty}
\Bigl ( 1-e^{2\pi i k\tau }\Bigr )
\Bigl ( 1-e^{2\pi i ((k-1)\tau +z)}\Bigr )
\Bigl ( 1-e^{2\pi i (k\tau -z)}\Bigr ).
\eeq
In order to unify some formulas given below, it is convenient to
understand the index $a$ modulo $4$, i.e., to identify
$\vartheta _{a} (z) \equiv \vartheta_{a+4} (z)$.
Set
$$
\omega _0 =0\,, \quad \omega_1 =\frac{1}{2}\,, \quad
\omega _2=\frac{1+\tau}{2}\,, \quad \omega _3 =\frac{\tau}{2}\,,
$$
then the function $\vartheta _a(z)$ has simple zeros at the points
of the lattice $\omega _{a-1}+\ZZ +\ZZ \tau $
(here $\omega_a \equiv \omega_{a+4}$).
The theta-functions have the following quasi-periodic properties
under shifts by $1$ and $\tau$:
\beq\label{Bp1a}
\begin{array}{l}
\vartheta _a (z+1)=e^{\pi i(1+\p_{\tau}\omega_{a\! -\! 1})}
\vartheta _a(z)
\\ \\
\vartheta _a (z+\tau )=e^{\pi i(a+\p_{\tau}\omega_{a\! -\! 1})}
e^{-\pi i\tau -2\pi iz}\vartheta _a(z).
\end{array}
\eeq
Shifts by the half-periods relate
the different theta-functions to each other:
\beq\label{shifts1}
\vartheta_1(z+ \omega_1)=\, \vartheta_2(z)\,,
\quad
\vartheta_3(z+ \omega_1)=\vartheta_0(z)\,,
\eeq
\beq\label{shifts2}
\vartheta_1(z+\omega_2)=e^{-\frac{\pi i \tau}{4}
-\pi i z} \vartheta_3(z)\,,
\quad
\vartheta_2(z+ \omega_2)=-ie^{-\frac{\pi i \tau}{4}
-\pi i z} \vartheta_0(z)
\eeq
\beq\label{shifts3}
\vartheta_1(z+ \omega_3)=ie^{-\frac{\pi i \tau}{4}
-\pi i z} \vartheta_0(z)\,,
\quad
\vartheta_2(z+ \omega_3)=e^{-\frac{\pi i \tau}{4}
-\pi i z} \vartheta_3(z).
\eeq

\paragraph{Weierstrass $\wp$-function.}
The Weierstrass $\wp$-function can be defined by the formula
\beq\label{Bp3}
\wp (z)= -\p_z^2 \log \vartheta _1 (z)-2\eta \,,
\eeq
where
\beq\label{Bp4}
\eta = -\, \frac{1}{6}\,
\frac{\vartheta _{1}^{'''}(0)}{\vartheta _1' (0)}=
-\, \frac{2\pi i}{3}\, \p_{\tau} \log \theta _1'(0|\tau ).
\eeq
The function $\wp (z)$ is double-periodic with periods
$2\omega _1 =1$, $2\omega _3 =\tau$,
$\wp (z+M+N\tau )=\wp (z)$, $M,N \in \ZZ$, and
has second order poles at the origin (and at all the
points $M +N\tau$ with integer $M$, $N$).
The derivative of the $\wp$-function is given by
\beq\label{Bp5}
\wp '(z)=-\,
\frac{2\, (\vartheta _1'(0))^3}{\vartheta_2(0)\vartheta_3(0)
\vartheta_0(0)}\,
\frac{\vartheta _2(z)\vartheta _3(z)
\vartheta _0(z)}{\vartheta _{1}^{3}(z)}\,.
\eeq

The values of the $\wp$-function at the half-periods,
$\omega_k$,
\beq\label{Bp6}
e_1 =\wp (\omega _1), \quad e_1 =\wp (\omega _2),
\quad e_3 =\wp (\omega _3)
\eeq
play a special role.
The sum of the numbers $e_k$ is zero: $e_1 +e_2 +e_3=0$.
The differences
$e_j-e_k$ can be represented
in terms of the values of the theta-functions at $z=0$
(theta-constants) in two different ways:
\beq\label{Bp7}
\begin{array}{l}
\displaystyle{e_1 -e_2 =\pi ^2 \vartheta_0^4 (0)\, =\,
4\pi i \, \p_{\tau} \log\frac{\vartheta_3 (0)}{\vartheta_2 (0)}}
\\ \\
\displaystyle{e_1 -e_3 =\pi ^2 \vartheta_3^4 (0)\, =\,
4\pi i \, \p_{\tau} \log\frac{\vartheta_0 (0)}{\vartheta_2 (0)}}
\\ \\
\displaystyle{e_2 -e_3 =\pi ^2 \vartheta_2^4 (0)\, =\,
4\pi i \, \p_{\tau} \log\frac{\vartheta_0 (0)}{\vartheta_3 (0)}}\,.
\end{array}
\eeq
The second representation is a consequence of the heat equation
(\ref{Bp2}) (see below). Its another consequence
is a representation of the $e_k$'s
themselves as logarithmic
$\tau$-derivatives of the theta-constants:
\beq\label{Bp8}
e_k = 4\pi i \, \p_{\tau}\Bigl (
\frac{1}{3}\, \log \vartheta _1'(0)
-\log \vartheta_{k+1}(0)\Bigr ).
\eeq
Using the first equalities in (\ref{Bp7}) and
the heat equation, the $\tau$-derivatives of the differences $e_j -e_k$
can be expressed through the $e_k$'s and $\eta$ as follows:
\beq\label{Bp9}
\pi i \, \p_{\tau}\log (e_j -e_k)=-e_l -2\eta \,.
\eeq
Here $\{jkl\}$ stands for any cyclic permutation of $\{123\}$.
Subtracting two such equations, we also get
\beq\label{Bp9a}
\pi i \, \p_{\tau}\log \frac{e_j -e_k}{e_l-e_k}=e_j-e_l \,.
\eeq

The $\wp$-function obeys the differential equation
\beq\label{Bp10}
(\wp '(z))^2 =4 (\wp (z)-e_1)(\wp (z)-e_2)(\wp (z)-e_3).
\eeq
We also mention the formulae
\beq\label{Bp11a}
\wp (z)-e_k =\frac{(\vartheta _1'(0))^2}{\vartheta_{k+1}^2(0)}\,
\frac{\vartheta_{k+1}^2(z)}{\vartheta_1^2(z)}\,.
\eeq

\paragraph{Eisenstein functions and $\Phi$-function.}

Sometimes it is convenient to use the Eisenstein functions
\beq\label{A.1}
E_1(z)=\p_z\log\vth(z)\,, \quad \quad
E_2(z)=-\p_zE_1(z)= -\p_z^2\log\vth(z)=\wp(z)+2\eta \,.
\eeq
The function $E_1$ is quasi-periodic,
$E_1(z+1)=E_1(z)$, $E_1(z+\tau)=E_1(z)-2\pi i$, while
$E_2$ is double-periodic: $E_2(z+1)=E_2(z)$, $E_2(z+\tau)=E_2(z)$.
Near $z=0$ they have the expansions
$$
E_1(z)=\frac{1}{z}-2\eta z  + \ldots \,,\quad \quad
~~E_2(z) =\frac{1}{z^2}+2\eta +\ldots
$$
It is not difficult to see that the function $E_1(z)$ has the
following values at the half-periods:
\beq\label{a21} E_1(\om_j)=-2\pi
i\p_\tau\om_j
\eeq
and, therefore, the identity
\beq\label{a22}
E_1(\om_j)+E_1(\om_k)=E_1(\om_j+\om_k)
\eeq
holds true for any
different $j,k =1,2,3$.

The following
function appears to be useful in the calculations:
\beq\label{Bp11}
\Phi(u,z)= \frac{\vth(u+z)\vth'(0)}{\vth(u)\vth(z)}\,.
\eeq
It obeys the obvious properties $\Phi(u,z)=\Phi(z,u)$,
$\Phi(-u,-z)=-\Phi(u,z)$
as well as less obvious ones:
\beq\label{Bp1101}
\Phi(u,z)\Phi(-u,z)=\wp(z)-\wp(u)
\eeq
\beq\label{Bp1102}
\Phi(u,z)\Phi(w,z)=\Phi(u+w,z)(E_1(z)+E_1(u)+E_1(w)-E_1(z+u+w)).
\eeq
Here $E_1(z)$ is the first Eisenstein
function. The expansion of the function $\Phi (u,z)$
near $z=0$ is
\beq\label{A.3a}
\Phi(u,z)=\frac{1}{z}+E_1(u)+\frac{z}{2}(E_1^2(u)-\wp(u))+
O(z^2).
\eeq
The quasi-periodicity properties of the
function $\Phi$ are:
\beq\label{A.14}
\Phi(u,z+1)=\Phi(u,z)\,,~~~\Phi(u,z+\tau)=e^{-2\pi i u}\Phi(u,z)\,.
\eeq
The $z$-derivative of the function $\Phi$ is equal to
\beq\label{A3b} \p_z \Phi (u,z)=\Phi(u,z)
(E_1(u+z)-E_1(z)).
\eeq

Finally, let us introduce the functions
\beq\label{Bp23}
\vf_j(z)=e^{2\pi i z\p_\tau \om_j}\Phi(z,\om_j)\,, \quad
j=1,2,3.
\eeq
Setting $u$ in (\ref{Bp1101}) and (\ref{Bp1102})
to be equal to the half-periods, we have:
\beq\label{Bp24}
\vf_j^2(z)=\wp(z)-e_j,\ \ \
\vf_j^2(z)-\vf_k^2(z)=e_k-e_j
\eeq
\beq\label{Bp25}
\vf_j(z)\vf_k(z)=\vf_l(z)(E_1(z)+E_1(\om_l)-E_1(z+\om_l)).
\eeq
In a similar way, from (\ref{A3b}) and (\ref{a21}) it follows that
\beq\label{Bp26}
\p_z\vf_j(z)=\vf_j (z)\Bigl [
E_1(z+\om_j )-E_1(\om_j)-E_1(z)\Bigr ]=-\vf_k(z)\vf_l(z),
\eeq
where $j,k,l$ is any cyclic permutation of $1,2,3$.

\subsection*{Heat equation and related formulae}

As it can be easily seen
from the definition (\ref{Bp1}),
all the theta-functions satisfy the ``heat equation''
\beq\label{Bp2}
4\pi i \p_{\tau}\vartheta _a(z|\tau )= \p_{z}^{2}\vartheta _a(z|\tau )
\eeq or, in terms of the variable
$\displaystyle{t=\frac{\tau}{2\pi i}}\, $ used in the main text,
$2\p_{t}\vartheta _a(z )=\p_{z}^{2}\vartheta _a(z)$.
One can also introduce the ``heat coefficient''
$\displaystyle{\kappa=\frac{1}{2\pi i}}$ and rewrite
the heat equation in the form
$\displaystyle{\p_{\tau}\vartheta _a(z|\tau )= \frac{\kappa}{2}\,
\p_{z}^{2}\vartheta _a(z|\tau )}$. All formulas for derivatives
of elliptic functions with respect to the modular parameter
are based on the heat equation.

The $\tau$-derivatives of the functions $\Phi$, $E_1$ and $E_2$
are given by the following proposition.
\begin{predl}
The identities
\beq\label{Bp12}
\p_\tau\Phi(z,u)=\kappa\p_z\p_u\Phi(z,u),
\eeq \beq\label{Bp13}
\p_\tau E_1(z)=\frac{\kappa}{2}\, \p_z(E_1^2(z)-\wp(z)),
\eeq
\beq\label{Bp14} \p_\tau E_2(z)=\kappa E_1(z)E_2'(z)-\kappa
E_2^2(z)+\frac{\kappa}{2}\, \wp''(z),
\eeq
with the ``heat coefficient'' $\displaystyle{\kappa=\frac{1}{2\pi i}}$,
hold true\footnote{(\ref{Bp12}) was obtained in
\cite{LO97},\cite{Takasaki02}.}.
\end{predl}
\underline{\emph{Proof}:}
First we prove (\ref{Bp12}). It follows from (\ref{Bp2}) that
\beq\label{Bp16}
4\pi i\frac{\p_\tau
\vth(z)}{\vth(z)}=\frac{\vth''(z)}{\vth(z)}=
\p_z\left(\frac{\vth'(z)}{\vth(z)}
\right)+\left(\frac{\vth'(z)}{\vth(z)}\right)^2=
-E_2(z)+E_1^2(z).
\eeq
Therefore,
\beq\label{Bp17}
\p_\tau \Phi(z,u)=
\frac{\kappa}{2}\left(-6\eta
\! -\! E_2(z+u)+E_1^2(z+u)+E_2(z)\! -\! E_1^2(z)+E_2(u)\! -\!
E_1^2(u)\right ),
\eeq
where the constant $\eta$ is given by (\ref{Bp4}).
On the other hand,
\beq\label{Bp18}
\begin{array}{lll}
\p_z\p_u\Phi(z,u)&=&\p_z \Bigl [\Phi(z,u)(E_1(z+u)-E_1(u))\Bigr
]=\Phi(z,u)
\\ &&\\
&\times&\Bigl [(E_1(z+u)-E_1(u))(E_1(z+u)-E_1(z))-E_2(z+u)\Bigr].
\end{array}
\eeq
The rest of the proof is a direct use of the identity
\beq\label{Bp19}
\left(E_1(z+u)-E_1(u)-E_1(z)\right)^2=\wp(z)+\wp(u)+\wp(z+u).
\eeq
Equation (\ref{Bp13}) easily follows from (\ref{Bp12}) and the local
expansion (\ref{A.3a}) around $u=0$. Equation (\ref{Bp14})
is just a derivative of (\ref{Bp13}).
\square

Next let us prove (\ref{6p9})\footnote{This formula was proved by
K.Takasaki in \cite{Takasaki01} by comparison of analytic properties
of the both sides. Here we give another proof by a direct computation.}.
\begin{predl}
Set $\displaystyle{X(z)=\frac{\wp(z)-e_1}{e_2-e_1}}$, then
\beq\label{Bp27}
\p_\tau X=\kappa \, \p_z X\,
\p_z \log \theta_0(z).
\eeq
\end{predl}
\underline{\emph{Proof}:} The $\tau$-derivative of
$\displaystyle{
X(z)=\frac{\wp(z)-e_1}{e_2-e_1}\stackrel{(\ref{Bp24})}{=}
\frac{\vf_1^2(z)}{e_2-e_1}}$ is:
$$
\p_\tau X=\frac{2\vf_1(z)\p_\tau\vf_1(z)
(e_2-e_1)-\p_\tau(e_2-e_1) \vf_1^2(z)}{(e_2-e_1)^2}\,.
$$
Using the definition of $\varphi_1(z)$ and the ``heat equation''
(\ref{Bp12}) for the $\Phi$-function, we write
\beq\label{Bp30}
\begin{array}{lll}
\p_\tau\vf_1(z)&
\stackrel{(\ref{A3b})}{=}&\kappa
\p_z\Bigl [\vf_1(z)(E_1(z+\om_1)-E_1(\om_1))\Bigr ]
\\ && \\
&=&
\kappa\p_z\Bigl [\vf_1(z)E_1(z+\om_1)\Bigr ]
\\ && \\
&=&
\kappa\p_z\vf_1(z) E_1(z+\om_1)-
\kappa \vf_1(z)E_2(z+\om_1).
\end{array}
\eeq
Substituting this and $\p_{\tau}(e_2-e_1)
\stackrel{(\ref{Bp9})}{=}-2\kappa (e_2-e_1)E_2(e_3)$
into (\ref{Bp30}), we have:
\beq\label{Bp31}
\p_\tau
X=\frac{2\kappa}{e_2-e_1}\Bigl (\vf_1(z)\p_z\vf_1(z)
E_1(z+\om_1)-\vf_1^2(z)E_2(z+\om_1)+E_2(\om_3)\vf_1^2(z)\Bigr ).
\eeq
Since
$\displaystyle{\p _z X=
\frac{2\vf_1(z)\p_z\vf_1(z)}{e_2-e_1}}$, we can rewrite the latter
equation as
$$
\p_\tau
X=\kappa \p_z X \, E_1(z+\om_1)+\frac{2\kappa
\vf_1^2(z)}{e_2-e_1}(-E_2(z+
\om_1)+E_2(\om_3))
$$
which can be further simplified with the help of the identity
$$
E_2(z+\om_1)=E_2(\om_1)+
\frac{(e_2-e_1)(e_3-e_1)}{\vf_1^2(z)}\,.
$$
Dividing both sides by $\p_z X$, we get
\beq\label{Bp34}
\frac{\p_{\tau} X}{\p_z X}=\kappa E_1(z+\om_1)+2\kappa
(e_3-e_1)\frac{X-1}{\p_z X}\,.
\eeq
The last term can be transformed using the
identities
$e_3-e_1\stackrel{(\ref{Bp24})}{=}\vf_1^2(z)-\vf_3^2(z)$,
$\displaystyle{\p_z X\stackrel{(\ref{Bp26})}{=}
-2\frac{\vf_1(z)\vf_2(z)\vf_3(z)}{e_2-e_1}}$
and $X-1\displaystyle{\stackrel{(\ref{Bp24})}{=}\frac{\vf_2^2(z)}{e_2-e_1}}$:
\beq\label{Bp35}
\p_\tau X=\kappa \p_z
X\left(E_1(z+\om_1)+\frac{\vf_2(z)\vf_3(z)}{\vf_1(z)}-\frac{\vf_1(z)
\vf_2(z)}{\vf_3(z)}\right).
\eeq
Finally, the desired formula (\ref{Bp27})
is obtained from this using (\ref{Bp25}):
\beq\label{Bp36}
\p_\tau
X=\kappa \p_z X\left(E_1(z+\om_3)-E_1(\om_3)\right)=\kappa \p_z X\p_z \log
\theta_0(z).
\eeq
\square

\section*{Appendix C}
\addcontentsline{toc}{section}{Appendix C}
\def\theequation{C\arabic{equation}}
\setcounter{equation}{0}

\subsection*{Gauge transformation of the linear
problems for ${\rm P}_{\rm VI}$}

In the parametrization (\ref{P6JM5}), (\ref{P6JM5a}),
the upper right entries of the matrices ${\sf U}(X,T)$,
${\sf V}(X,T)$ forming the modified Jimbo-Miwa $U$--$V$ pair
for the ${\rm P}_{\rm VI}$ equation are
$$
{\sf U}_{12}=
{\sf b}=\frac{K(X-y)}{X(X-1)(X-T)}\,,
\quad \quad
{\sf V}_{12}=
{\sf B}=\frac{K(y-T)}{T(T-1)(X-T)}\,.
$$
Passing to a parametrization
$X=X(x,t)$, $T=T(t)$ according to the rule
(\ref{6p3}) and performing a diagonal gauge transformation
of the form (\ref{tilde1}) we get the following expressions
for the upper right entries of the matrices ${\bf U}(x,t)$,
${\bf V}(x,t)$:
\beq\label{s2} b={\bf U}_{12}
={\sf b}X_x\om^2=\frac{K(X-y)}{X(X-1)(X-T)}X_x\om^2 \,,
\eeq
\beq\label{s3}
B={\bf V}_{12}=(T_t {\sf B}+X_t {\sf b})\om^2=
\frac{K(y-T)}{T(T-1)(X-T)}
T_\tau \om^2+\frac{K(X-y)}{X(X-1)(X-T)}X_t \om^2 \,.
\eeq
The $x$-derivative of $b$ is
\beq\label{s4}
b_x= \frac{(X-y)\om^2}{X(X-1)
(X-T)}\left (  f+ \frac{ X_x^2}{X-y} \right ),
\eeq
where the notation
$$
f=X_{xx} +X_x \p_x\log(\om^2)- X_x^2
\left(\frac{1}{X}+\frac{1}{X-1}+\frac{1}{X-T}\right)
$$
is introduced for brevity.
Further, let us impose condition of the form (\ref{Bb}):
\beq\label{s5}
b_x=k B,
\eeq
with some constant $k$ (not yet fixed).
Substituting (\ref{s3}) and
(\ref{s4}) into (\ref{s5}), we obtain an equality of two
linear functions of $y$ {\it provided $\p_x \log \omega$
does not depend on $y$}. (The latter assumption is necessary
to achieve separation of the variables $x$, $u$ in the
non-stationary Schr\"odinger equation.)
Assuming this, we equate the coefficients
in front of $y$ and the $y$-independent terms
in the both sides and get the system of two equations
\beq\label{s7}
\left\{\begin{array}{l}\displaystyle{f=
k\left ( X_t -\frac{X(X-1)}{T(T-1)}\, T_t \right)}
\\ \\
\displaystyle{
X f+\left(X_x \right)^2=k X
\left (X_t -\frac{X-1}{T-1}\, T_t \right ).}
\end{array}\right.
\eeq
from which the functions $X(x,t)$ and $\p_x \log \omega$
can be determined.
Excluding $f$, we arrive at the differential equation for
$X$:
\beq\label{s8}
X_x^2=\frac{kT_t}{T(T-1)}X(X-1)(X-T).
\eeq
We know that $T_t$ is given by
(\ref{6p8}): $T_t=2(e_2-e_1)T(T-1)$. Therefore,
\beq\label{s9}
X_x^2=2k (e_2-e_1)X(X-1)(X-T).
\eeq
This relation prompts the elliptic parametrization
(\ref{6p1}) and fixes the value of $k$:
\beq\label{s11}
k=2\,.
\eeq
Note that in some sense this is ``the same'' coefficient
$2$ that enters the heat equation for theta-functions
in the $t$-variable $t=\kappa \tau$: $2\p_t \vartheta _a(x)=
\p_{x}^{2}\vartheta _a(x)$. In the same sense
the non-stationary Schr\"odinger equation for
the $\psi$-function is a ``dressed" version of the
heat equation.

Now we are ready to fix the $x$-dependent part of
the function
$\om^2$. From the first equation of the system (\ref{s7}) we find:
\beq\label{s12}
\p_x\log\om^2=-\frac{X_{xx}}{X_x}+X_x \left (\frac{1}{X}+
\frac{1}{X-1}+\frac{1}{X-T}\right)
-k\frac{X(X-1)}{T(T-1)}\frac{T_t}{X_x}+k\frac{X_t}{X_x}\,.
\eeq
It is easy to show that
$\displaystyle{
\frac{X_{xx}}{X_x}=\frac{X_x}{2}\left(\frac{1}{X}+\frac{1}{X-1}+
\frac{1}{X-T}\right)}$, so plugging the previously obtained
formulas for $X_t$ and $T_t$ into (\ref{s12}), we get:
\beq\label{s13}
\p_x\log\om^2 \! =\!
\frac{X_x}{2}\left(\frac{1}{X}+\! \frac{1}{X-1}+\! \frac{1}{X\! -\! T}\right)
-4(e_2 -\! e_1)\frac{X(X\! -\! 1)}{X_x}+
2E_1(x\! +\! \om_3)-2E_1(\om_3).
\eeq
To proceed, we substitute
$$
X=\frac{\vf_1^2(x)}{e_2-e_1},\quad X-1=\frac{\vf_2^2(x)}{e_2-e_1},\quad
X-T=\frac{\vf_3^2(x)}{e_2-e_1}
$$
and
$$
X_x=-2\, \frac{\vf_1(x)\vf_2(x)\vf_3(x)}{e_2-e_1}\,.
$$
This yields
\beq\label{s14}
\p_x\log\om^2=-\frac{\vf_2(z)\vf_3(z)}{\vf_1(z)}-
\frac{\vf_1(z)\vf_3(z)}{\vf_2(z)}+
\frac{\vf_1(z)\vf_2(z)}{\vf_3(z)} +2E_1(x+\om_3)-2E_1(\om_3).
\eeq
The final result obtained with the help of (\ref{Bp25}) is
\beq\label{s15}
\p_x\log\om^2=-E_1(x)+\sum_{j=1}^{3}\Bigl [
E_1(x+\om_j)-E_1(\om_j)\Bigr ],
\eeq
or, in the integrated form,
\beq\label{s16}
\om^2(x,t)=\frac{\vartheta_2(x)\vartheta_3(x)
\vartheta_0(x)}{\vth(x)}\, g(y,t),
\eeq
where the function $g(y,t)$ can not be fixed by
the above arguments. Using the identity
$$
2\, \frac{\vth '(0)\vartheta_0(0)}{\vartheta_2(0)\vartheta_3(0)}\,
\frac{\vartheta _2(x)\vartheta _3(x)}{\vth (x)\vartheta _0(x)}
=-\, \frac{\wp '(x)}{\wp (x)-e_3}\,,
$$
we can express $\omega^2$ in terms of the $\wp$-function:
\beq\label{s17}
\omega^2 (x,t)=\frac{\wp '(x)\theta _0^2(x)}{2(\wp (x)-e_3)}
\, \rho ^2(t)
\eeq
with some $\rho (t)$ to be determined from the condition that the
$x$-independent part of the potential in the non-stationary
Schr\"odinger equation be equal to the classical
Hamiltonian $H_{\rm VI}(\dot u, u)$. It is the form (\ref{s17})
that is more convenient to use in Section 8.3.


\begin{thebibliography}{99}
\addcontentsline{toc}{section}{References}

\bibitem{Painleve1} P.Painlev\'e, {\it Memoire sur les
\'equations diff\'erentielles dont l'int\'egrale g\'en\'erale
est uniforme}, Bull. Soc. Math. Phys. France {\bf 28} (1900)
201-261;\\
P.Painlev\'e, {\it Sur les \'equations diff\'erentielles du
second ordre et d'ordre sup\'erieur dont l'int\'egrale g\'en\'erale
est uniforme}, Acta Math. {\bf 21} (1902) 1-85

\bibitem{Fuchs} R.Fuchs, {\it Sur quelques
\'equations diff\'erentielles lin\'eares du second ordre},
C. R. Acad. Sci. (Paris) {\bf 141} (1905) 555-558

\bibitem{Gambier} B.Gambier, {\it Sur les \'equations diff\'erentielles
du second ordre et du premier degr\'e dont l'int\'egrale g\'en\'erale
est \`a points critique fix\'es},
C. R. Acad. Sci. (Paris) {\bf 142} (1906) 266-269


\bibitem{book} K.Iwasaki, H.Kimura, S.Shimomura, M.Yoshida,
{\it From Gauss to Painlev\'e, a modern theory of special
funtions}, Aspects of Mathematics, {\bf E16},
Friedr. Vieweg \& Sohn, Braunschweig, 1991

\bibitem{book1}
{\it The Painleve Property. One Century Later},
CRM Series in Mathematical Physics, XXVI, R.Conte (Ed.),
1999, 810 p.

\bibitem{FN} H.Flaschka and A.Newell, {\it Monodromy- and
spectrum-preserving deformations. I}
Commun. Math. Phys. {\bf 76} (1980) 65-116

\bibitem{Bar} E.Barouch, B.McCoy, C.Tracy and T.Wu,
{\it Zero field susceptibility of the two-dimensional Ising
model near $T_c$}, Phys. Rev. Lett. {\bf 31} (1973) 1409-1411

\bibitem{JMMS} M.Jimbo, T.Miwa, Y.Mori and M.Sato,
{\it Density matrix of an impenetrable gas and the fifth
Painlev\'e transcendent}, Physica {\bf D1} (1980) 80-158

\bibitem{qg} E.Br\'ezin and V.Kazakov, {\it Exactly
solvable field theories of closed strings},
Phys. Lett. {\bf B236} (1990) 144-150;\\
D.Gross and A.Migdal, {\it Nonperturbative two-dimensional
quantum gravity}, Phys. Rev. Lett. {\bf 64} (1990) 127-130;\\
M.Douglas and S.Shenker, {\it Strings in less than one dimension},
Nuclear Physics {\bf B335} (1990) 635-654

\bibitem{Zamolodchikov} Al.Zamolodchikov,
{\it Painlev\'e III and 2D polymers}, Nuclear Physics
{\bf B432} (1994) 427-456

\bibitem{TW} C.Tracy and H.Widom, {\it Fredholm determinants,
differential equations and matrix models},
Commun. Math. Phys. {\bf 163} (1994) 33-72

\bibitem{FW} P.Forrester and N.Witte, {\it
Application of the $\tau$-function theory of Painlev\'e
equations to random matrices: ${\rm PIV}$, ${\rm PII}$
and the GUE}, Commun. Math. Phys. {\bf 219} (2001) 357-398;\\
P.Forrester and N.Witte, {\it
Random matrix theory and the sixth Painlev\'e equation},
J. Phys. A: Math. Gen. {\bf 39} (2006) 12211-12233

\bibitem{Dubrovin}
B.Dubrovin, {\it Geometry of 2D topological field theories},
Integrable systems and quantum
groups (Montecatini Terme, 1993),
Lecture Notes in Math., vol. 1620,
Springer, Berlin 1996, pp. 120-348;\\
B.Dubrovin, {\it Painlev\'e equations in 2D topological field theories},
In: Painleve Property, One Century Later, Carg\'ese, 1996,
arXiv:math.AG/9803107

\bibitem{LTW} S.-Y.Lee, R.Teodorescu and P.Wiegmann,
{\it Viscous shocks in Hele-Shaw flow and
Stokes phenomena of the Painleve I transcendent}, Physica
{\bf D240} (2011) 1080-1091



\bibitem{Garnier}R.Garnier,
{\it Sur des equations diff\'erentielles du troisi\'eme ordre
dont l'int\'egrale g\'en\'erale est uniforme et sur une classe
d'\'equations nouvelles d'ordre sup\'erieur dont l'int\'egrale
g\'en\'erale a ses points critique fix\'es},
Ann. Ecol. Norm. Sup. {\bf 29} (1912) 1-126

\bibitem{Schlesinger} L.Schlesinger, {\it \"Uber eine Klasse
von Differentialsystemen beliebiger Ordnung mit feten kritischen
Punkten}, J. Reine u. Angew. Math. {\bf 141} (1912) 96-145



\bibitem{JM81-1} M.Jimbo, T.Miwa and K.Ueno, {\it Monodromy preserving
deformation of linear ordinary differential equations with
rational coefficients I. General theory
and $\tau$-function}, Physica D {\bf 2} (1981) 306-352

\bibitem{JM81-2} M.Jimbo and T.Miwa, {\it Monodromy preserving
deformation of linear ordinary differential equations with
rational coefficients II}, Physica D {\bf 2} (1981) 407-448

\bibitem{JM81-3} M.Jimbo and T.Miwa, {\it Monodromy preserving
deformation of linear ordinary differential equations with
rational coefficients III}, Physica D {\bf 4} (1981) 26-46


\bibitem{IN} A.Its and V.Novokshenov, {\it The isomonodromic
deformation method in the theory of Painlev\'e equations},
Lecture Notes in Math. {\bf 1191} (1986), Berlin: Springer;\\
A.Fokas,  A.Its, A.Kapaev and V.Novokshenov,
{\it Painlev\'e transcendents: the Riemann-Hilbert approach},
AMS Mathematical Surveys and Monographs, vol. 128,
Providence, RI, 2006








\bibitem{JKT07} N.Joshi, A.Kitaev and P.Treharne,
{\it On the linearization of the Painlev\'e III-VI equations
and reductions of the three-wave resonant system},
J. Math. Phys. {\bf 48} (2007) 103512 (42 pages),
arXiv:0706.1750


\bibitem{Malmquist} J.Malmquist, {\it Sur les \'equations
diff\'erentielles du second ordre dont l'int\'egrale g\'en\'erale
a ses points critique fixes}, Ark. Mat. Astr. Fys.
{\bf 17} (1922/23) 1-89

\bibitem{DM07}
B.Dubrovin and M.Mazzocco, {\it Canonical structure
and symmetries of the Schlesinger
equations}, Commun. Math. Phys. {\bf 271} (2007) 289-373



\bibitem{Okamoto} K.Okamoto,
{\it On the $\tau$-function of the Painlev\'e
equations}, Physica D {\bf 2} (1981) 525-535;\\
K.Okamoto, {\it Isomonodromic deformations and
Painlev\'e equations, and the Garnier systems},
J. Fac. Sci. Univ. Tokyo, Sect. IA Math. {\bf 33}
(1986) 575-618;\\
K.Okamoto, {\it Polynomial Hamiltonians associated with
Painlev\'e equations. I}, Proc. Japan Acad. Ser. A {\bf 56}
(1980) 264-268


\bibitem{LO97} A.Levin and M.Olshanetsky,
{\it Painlev\'e-Calogero correspondence},
Calogero-Moser-Sutherlend models (Montreal, 1997),
CRM Ser. Math. Phys., Springer 2000, pp. 313--332,
arXiv: alg-geom/9706010.

\bibitem{Inoz} V.I.Inozemtsev and D.V.Meshcheryakov,
{\it Extension of the class of integrable
dynamical systems connected with semisimple Lie algebras},
Lett. Math. Phys. {\bf 9} (1985) 13-18;\\
V.I.Inozemtsev, {\it Lax representation with spectral
parameter on a torus for integrable particle systems},
Lett. Math. Phys. {\bf 17} (1989) 11-17.

\bibitem{Manin98} Yu.Manin, {\it Sixth Painlev\'e equation,
universal elliptic curve, and mirror of $\PP ^2$},
AMS Transl. (2) {\bf 186} (1998) 131-151

\bibitem{Painleve1906} P.Painlev\'e, {\it Sur les
\'equations diff\'erentielles du second ordre \`a
points critiques fix\'es}, C. R. Acad. Sci. (Paris)
{\bf 143} (1906) 1111-1117


\bibitem{Takasaki01}
K.Takasaki, {\it Painlev\'e-Calogero correspondence revisited},
J. Math. Phys. {\bf 42} (2001) 1443-1473

\bibitem{Suleimanov94}
B.Suleimanov, {\it The Hamiltonian property of
Painlev\'e equations and the method of isomonodromic deformations},
Differential Equations {\bf 30:5} (1994) 726-732
(Translated from Differentsialnie
Uravneniya {\bf 30:5} (1994) 791-796)

\bibitem{Suleimanov08}
B.Suleimanov, {\it ``Quantizations'' of the second Painlev\'e
equation and the problem of the equivalence of its $L$-$A$ pairs},
Theor. Math. Phys. {\bf 156} (2008) 1280-1291
(Translated from Teor. Mat. Fys. {\bf 156} (2008) 364-377)

\bibitem{DNovikov09}
D.Novikov, {\it The 2$\times$2 matrix Schlesinger system and
the Belavin-Polyakov-Zamolodchikov system},
Theor. Math. Phys. {\bf 161} (2009) 1485-1496
(Translated from Teor. Mat. Fys. {\bf 161} (2009) 191-203)

\bibitem{VN84}
A.Veselov and S.Novikov, {\it Poisson brackets and complex tori},
Trudy Mat. Inst. Steklov, {\bf 165} (1984) 49-61

\bibitem{Sklyanin}
E.Sklyanin, {\it Separation of variables. New trends},
In: Quantum field theory, integrable models and beyond
(Kyoto, 1994), Progr. Theor. Phys. Suppl. {\bf 118}
(1995) 35-60



\bibitem{Takasaki02}
K.Takasaki, {\it Elliptic Calogero-Moser systems and isomonodromic
deformations}, J. Math. Phys. {\bf 40}, (1999) 57-87


\bibitem{Joshi06} P.Gordoa, N.Joshi and A.Pickering,
{\it Second and fourth Painlev\'e hierarchies and Jimbo-Miwa
linear problems}, J. Math. Phys. {\bf 47} (2006), pp. 073504

\bibitem{Babich} M.Babich,
{\it On canonical parametrization of the phase spaces of equations of
isomonodromic deformations of Fuchsian systems of dimension
$2 \times 2$.
Derivation of the Painlev\'e VI equation},
Russian Mathematical Surveys {\bf 64}:1 (2009) 45-127

\bibitem{Guzzetti}
D.Guzzetti, {\it The elliptic representation of the
general Painlev\'e VI equation},
Comm. Pure Appl. Math. {\bf 55}:10 (2002) 1280-1363


\bibitem{Krichever02} I.Krichever, {\it
Isomonodromy equations on algebraic curves,
canonical transformations and Whitham equations},
Moscow Math. J. {\bf 2} (2002) 717-806, arXiv:hep-th/0112096


\bibitem{Zotov04} A.Zotov,
{\it Elliptic linear problem for Calogero-Inozemtsev model and
Painlev\'e VI equation}, Lett. Math. Phys. {\bf 67} (2004) 153-165,
arXiv:hep-th/0310260

\bibitem{LZ06}
A.Levin and A.Zotov, {\it On rational and
elliptic forms of Painlev\'e VI equation},
Moscow Seminar on Mathematical Physics, II,
American Mathematical Society, Translations, Ser. 2,
Vol. 221, 173-184 (2007)

\bibitem{BM06} V.Bazhanov and V.Mangazeev, {\it
The eight-vertex model and Painlev\'e VI}, J. Phys. A:
Math. Gen. {\bf 39} (2006) 12235-12243

\bibitem{FL} V.Fateev and I.Litvinov, {\it On AGT conjecture},
JHEP {\bf 1002} (2010) 014, arXiv:0912.0504

\bibitem{MMM} A.Marshakov, A.Mironov and A.Morozov,
{\it On AGT relations with surface operator insertion
and stationary limit of beta-ensembles},
J. Geom. Phys. {\bf 61} (2011) 1203-1222

\bibitem{Reshet92} N.Reshetikhin, {\it The Knizhnik-Zamolodchikov
system as a deformation of the isomonodromy problem},
Lett. Math. Phys. {\bf 26} (1992) 167-177

\bibitem{Harnad94} J.Harnad, {\it
Quantum isomonodromic deformations and the
Knizhnik--Zamolodchikov equations},
CRM Proc. Lecture Notes {\bf 9} 155-161 (Amer. Math. Soc., Providence, RI, 1996), arXiv:hep-th/9406078

\bibitem{Slavyanov} S.Slavyanov, {\it Painlev\'e equations
as classical analogues of Heun equations}, J. Phys. A: Math. Gen.
{\bf 29} (1996) 7329-7335;\\
S.Slavyanov and W.Lay, {\it Special functions: a unified theory
based on singularities}, Oxford; New York:
Oxford University Press, 2000



\end{thebibliography}
\end{document}